\documentclass[12pt,a4paper]{article}
\usepackage[english]{babel}
\usepackage[letterpaper,top=2cm,bottom=2cm,left=3cm,right=3cm,marginparwidth=1.75cm]{geometry}
\usepackage[numbers]{natbib}
\usepackage{amsmath}
\usepackage{amssymb}
\usepackage{graphicx}
\usepackage{tikz}
\usetikzlibrary{calc,trees,positioning,arrows,chains,shapes.geometric,%
    decorations.pathreplacing,decorations.pathmorphing,shapes,%
    matrix,shapes.symbols,shadows}
\usepackage[font={small,it}]{caption}
\usepackage{subcaption}
\usepackage[colorlinks=true, allcolors=blue]{hyperref}
\usepackage{comment}
\usepackage{algorithm}
\usepackage{algpseudocode}
\usepackage{authblk}
\usepackage{indentfirst}
\usepackage{lineno, hyperref}
\modulolinenumbers[5]

\begin{document}
\title{Ricci flow-based brain surface covariance descriptors for diagnosing Alzheimer's disease}
\author[1]{Fatemeh Ahmadi}
\author[1]{Mohamad-Ebrahim Shiri \footnote{Corresponding author.\\Emails: f.ahmadi@aut.ac.ir (F. Ahmadi), shiri@aut.ac.ir (M. E. Shiri), bidabad@aut.ac.ir (B. Bidabad), m\_sedaghat@aut.ac.ir (M. Sedaghat), memari@lix.polytechnique.fr (P. Memari) }}
\author[1]{Behroz Bidabad}	
\author[1]{Maral Sedaghat}	
\author[2]{Pooran Memari}	
\affil[1]{\small Department of Mathematics and Computer Science, Amirkabir Unversity of Technology (Tehran Polytechnic), Tehran, Iran.}	
\affil[2]{LIX, CNRS - Ecole Polytechnique, IP Paris, Inria, France.}
\date{}
\maketitle
\abstract{Automated feature extraction from MRI brain scans and diagnosis of Alzheimer's disease are ongoing challenges. With advances in 3D imaging technology, 3D data acquisition is becoming more viable and efficient than its 2D counterpart. 
\textcolor{black}{Rather than using feature-based vectors, in this paper, for the first time, we suggest a pipeline to extract novel covariance-based descriptors from the cortical surface using the Ricci energy optimization.} The covariance descriptors are components of the nonlinear manifold of symmetric positive-definite matrices, thus we focus on using the Gaussian radial basis function to apply manifold-based classification to the 3D shape problem. Applying this novel signature to the analysis of abnormal cortical brain morphometry allows for diagnosing Alzheimer's disease. Experimental 
\textcolor{black}{studies performed on about two hundred 3D MRI brain models, gathered} from Alzheimer's Disease Neuroimaging Initiative (ADNI) dataset 
\textcolor{black}{demonstrate the effectiveness of our descriptors in achieving remarkable classification accuracy}.} 

\textbf{Keywords:}  Brain Mapping, Discrete Ricci Flow, Conformal Parameterization, Surface Classification, Covariance Matrix, Heat Kernel, Alzheimer's Disease.



\maketitle

\section{Introduction}\label{sec1}
Cognitive dysfunction that worsens with time is a feature of some neurodegenerative disorders, including Alzheimer's disease (AD). Medically, AD is distinguished by all-encompassing dementia symptoms such as executive disorder, forgetfulness, and impairment of language. It is most likely that the pathology of the underlying disease develops several years before the emergence of cognitive symptoms. Research is being done to identify early diagnostic biomarkers that will allow for a quick and accurate assessment of neurodegenerative risk prior to the onset of symptoms. Structural magnetic resonance imaging (MRI) assessments of brain shrinkage are some of the most well-established indicators of AD development and pathology among the various brain imaging techniques used for early identification and tracking of AD. 

Early MRI studies of brain structure showed that surface-based studies might have advantages over volume-based studies for examining the complexity and patterns of change over time brought on by illness or developmental processes \cite{van1998functional, dale1999cortical, ashburner1998identifying}. As a result, surface-based methods have recently drawn much interest and have been used in brain morphometry to explore anomalies \cite{shimony2016comparison, besson2021geometric}, such as spherical harmonic analysis \cite{chung2008encoding}, conformal invariants \cite{shi2017conformal}, brain gyrification index \cite{zakharova2021brain}, sulci pits, and patterns \cite{im2019sulcal}, and so on. These techniques can examine morphological changes or anomalies in cortical and subcortical brain regions. Unlike the majority of the existing methods concentrated on local geometric signatures, such as thickness or distance, our approach emphasizes intrinsic geometry using feature changes in Ricci energy optimization. Local angle-invariant descriptors, through steps in Ricci flow, are used to quantify the intrinsic conformal structure. 

Ricci flow is a powerful method for computing conformal structures on any arbitrary surface. It was used to demonstrate Poincaré's conjecture efficiently. Like heat diffusion, the Ricci flow alters the Riemannian metric to match the desired curvature, resulting in uniform curvatures across the surface.

The discrete surface Ricci flow has been introduced in \cite{wang2011brain, zhang2015survey, Jin2018Discrete}. Our technique computes a variety of features that are intrinsically related to the surface using Ricci flow. It is challenging to combine these distinct shape features because of their dissimilar dimensions, sizes, and ranges of variation. To address this, we combine diverse modalities via covariance matrices, generating unordered covariance descriptors through Ricci energy optimization for 3D shape indexing.

Two main properties of covariance matrices motivate our work. First, they efficiently combine multi-modal features without normalization nor high dimensional joint distribution estimates. Second, the covariance matrices produced are significantly more compact than the combined attributes and their statistical data. Their usage thus enables comparison of the subjects without being constrained to a particular feature dimension.

For classification based on covariance matrices, we use nonlinear mappings to Riemann manifolds or Hilbert space \cite{tuzel2008pedestrian,yun2016exploiting,khan2014online} by Gaussian radial basis functions to construct the vector spaces in which the metrics of the machine learning techniques are developed. This setting allows us to then simply employ the K-nearest neighbor method for classification purposes. The efficacy of this approach is demonstrated by applying it to the comparison of hippocampal regions in the cortical surface between Alzheimer's patients and cognitively normal individuals. This region specification is based on the fact that Alzheimer's disease, as a degenerative neurological illness that causes cell death and brain atrophy, highly affects the hippocampus, a region closely linked to memory functions \cite{sabuncu2011dynamics, thangavel2023ead}.

The following is a summary of the key contributions presented in the paper:
\begin{itemize}
\item A novel approach is introduced for diagnosing Alzheimer's disease utilizing the 3D brain cerebral cortex.
\item \textcolor{black}{The methodology entails extracting local features throughout the optimization steps of the Ricci flow. Intrinsic features, such as the conformal factor, area distortion, and heat kernel signatures that carry valuable information, are extracted. }
\item \textcolor{black}{By utilizing the covariance matrices, the extracted features are effectively combined. The covariance matrix provides a unified representation that captures the collective behavior of the features and allows for quick comparison and classification based on surface intrinsic properties.}
\item \textcolor{black}{Due to the nonlinearity of the SPD manifold, the covariance matrices on the Riemannian manifold are mapped into a higher-dimensional Hilbert space using the RBF kernel function to render the KNN classification algorithm feasible.}
\item To the best of our knowledge, the proposed method is the first to apply covariance descriptors for surface indexing using the Ricci flow.
\end{itemize}
\noindent The paper's organization is as follows: Section 2 provides a literature overview, Section 3 presents the mathematical background, Section 4 details the proposed pipeline, Section 5 provides experimental results, and we conclude in Section 6.
\section{Related work}\label{sec2}
In brain surface analysis, higher-order correspondences between particular anatomical locations, curved landmarks, or subregions located within the two surfaces are frequently created. This is accomplished by parameterizing the surface and embedding it in canonical parameter spaces. Surface parameterization establishes a mapping from a 2D surface embedded in $\mathbb{R}^3$ to $\mathbb{R}^2$, thus making it possible to unfold the object of interest so that it may be mapped onto more basic 2D structures, such as plane or sphere.

During the last decades, various surface parametrization techniques have been introduced \cite{floater2005surface,kreiser2018survey}. Some techniques parameterize and integrate the cortical surface directly into the sphere domain by optimizing certain energy functions \cite{nadeem2016spherical}. Other approaches cut the cortical surface and parametrize it planarly \cite{li2018optcuts}. The Ricci flow, as a parametrization technique, enables the conformal mapping of a surface to a canonical parameter space, such as a sphere, a Euclidean plane, or a hyperbolic plane \cite{Jin2018Discrete}. Compared to other conformal parametrization methods studied in \cite{wang2011brain,gu2004genus}, the Ricci flow method can handle cortical surfaces with complicated topology without producing singularities \cite{wang2011brain}. By the continuous Ricci flow, a Riemannian metric is conformally deformed over a smooth surface in a way that causes the Gaussian curvature to change like heat diffusion. The Newton technique can be used to formulate and solve the Ricci flow in the discrete framework using the circle packing metric \cite{Jin2018Discrete}.  

More recently, Ricci flow has also been used in brain surface morphometry to diagnose brain abnormalities and diseases. In \cite{chen2013ricci}, the authors have suggested a novel method for computing spherical parameterization based on the Euclidean Ricci flow through a different formulation of discrete Gaussian curvature. They used a scale space processing based on Ricci energy to extract the geometric property of the surface and apply it to the analysis of the region of the hippocampus in Schizophrenia disease. Zeng et al. \cite{zeng2013teichmuller} applied Ricci flow to calculate a Teichmüller shape descriptor that uses both global and local features to represent a closed surface of the cortical brain with genus zero. The method applied to analysis abnormalities in the hippocampal region, which is used to diagnose Alzheimer's disease. Shi et al. \cite{shi2017conformal} proposed to use the surface Ricci flow method to map a multi-connected surface to the Poincaré disk and computed a set of conformally invariant shape indices associated with the boundary lengths induced by the landmark curve in the hyperbolic parameter domain. They assessed the technique for analyzing the anomalies in brain morphometry linked to Alzheimer's disease. 

It is also notable that Ricci flow has also been applied to graphs to solve some machine learning problems, such as community detection in complex networks \cite{ni2019community}, classification, reduction, and visualization of high-dimensional data \cite{xu2023camel}, and measuring the structure of complex networks and the impact of objects on graph weights by comparing two graphs \cite{cohen2022object}.

\textcolor{black}{Given the importance of research in the diagnosis of Alzheimer's disease, numerous techniques of artificial neural networks to diagnose Alzheimer's disease have emerged recently. Lei et al. \citep{lei2023multi} presented a multi-scale enhanced graph convolutional network (MSE-GCN) for diagnosing mild cognitive impairment (MCI). The suggested technique combines the functional and structural data from resting-state functional magnetic resonance imaging (R-fMRI) and diffusion tensor imaging (DTI). In order to visualize the morphological features that indicate the severity of AD for patients in different stages, Yu et al. \citep{yu2022morphological} proposed a novel multidirectional perception generative adversarial network (MP-GAN). Zuo et al. \citep{zuo2021multimodal} used entire trimodal images to develop a novel multi-modal representation learning and adversarial hypergraph fusion (MRL-AHF) framework for diagnosing Alzheimer's disease.
In order to diagnose Alzheimer's disease with the class-imbalance problem, Hu et al. \citep{hu2020medical} devised an efficient generative adversarial network method for medical image reconstruction.
Pan et al. \citep{pan2021characterization} proposed a unique Hypergraph Generative Adversarial Networks (HGGAN) that generates multi-modal connectivity of Brain Network from rs-fMRI in combination with DTI by using Interactive Hyperedge Neurons module (IHEN) and Optimal Hypergraph Homomorphism method (OHGH).}

\textcolor{black}{While MRI-based methods are currently widely used to diagnose Alzheimer's disease, focusing on the study of the cerebral cortex offers advantages which may improve diagnosis in combination with other methods.} To analyze the cortical surface for disease diagnosis, methods typically use distinct landmarks to guide the diagnostic procedure for cortical surface analysis in pathological diagnosis. 
Although processing using manual landmarks is superior to processing with common surface characteristics, it is challenging to apply to large volumes of data and requires medical expertise. Therefore, it is essential to concentrate on the landmark-free method and local processing of multiple surface properties for cerebral cortex morphometry, as we perform in the current study. Since Alzheimer's disease more severely affects the hippocampus region than other regions, it is more straightforward and efficient to examine a specific region rather than the entire brain. In this study, we use the Euclidean Ricci flow method to compute a planar conformal parameterization. We focus on local processing and a landmark-free approach using different surface properties in Ricci energy optimization. We propose merging these multi-modal properties using covariance matrices. In the literature, covariance matrices have been successfully used for image set classification \cite{wang2012covariance,fang2018new}, face recognition \cite{hariri20173d}, and shape retrieval \cite{tabia2015covariance}. In our approach, a set of unordered covariance descriptors computed through Ricci energy optimization are used to index a surface. Through the RBF kernel, these covariance matrices are mapped into a high-dimensional Hilbert space, enabling us to apply traditional classification methods to non-linear data.
\section{Mathematical background}
\subsection{Ricci flow algorithm}
Given the triangular mesh $M$ with the initial Euclidean metric and the given target curvature $\bar{K}: V \rightarrow \mathbb{R}$, where $V$ is the vertex set of the mesh. The goal is to calculate a discrete Riemannian metric that is conformal with the original metric and produces the desired curvature. To solve this issue, we will review the formulas for calculating the Ricci flow on discrete surfaces with an Euclidean background geometry \cite{Jin2018Discrete}. 

First, the initial circle packing metric is calculated using the inversive distance circle packing scheme, as outlined in Algorithm \ref{algorithm1}. For each vertex $ v_i$ surrounded by faces $f_{ijk}$, we determine the circle's radius for the vertex $ v_i$ using edge length as
\begin{equation}
	\gamma_i^{jk} = \dfrac{l_{ij}+l_{ki} - l_{jk}}{2}.
\end{equation}
Given the Euclidean geometry of the mesh background, for each edge $e_{ij} $, the inverse distance between two circles $(v_i,\gamma_i)$ and $(v_j,\gamma_j)$ is defined as 
\begin{equation}
	\eta_{ij}=\dfrac{l_{ij}^2-\gamma_i^2-\gamma_j^2}{2\gamma_i\gamma_j}.
\end{equation}

\begin{algorithm}[!h] \fontsize{10pt}{10pt}\selectfont
\caption{Primary circle packing metric}\label{algorithm1}
\textbf{Input:} $M \in \mathbb{R}^3$: a triangular mesh.\\
\textbf{Output:} An initial circle packing metric.
\begin{enumerate}
    \item \textbf{for each face} $f_{ijk} \in M$ \textbf{do}
    \item \quad \quad Calculate $ \gamma_i^{jk} = \dfrac{l_{ij}+l_{ki} - l_{jk}}{2} $.
    \item \textbf{end for}
    \item \textbf{for each vertex} $ v_i \in M $ \textbf{do}
    \item \quad \quad Calculate the radius $ \gamma_i = min_{jk} \gamma_i^{jk}$. 
    \item \textbf{end for}
    \item \textbf{for each edge} $e_{ij} \in M $ \textbf{do}
    \item \quad \quad Calculate the inversive distance $ \eta_{ij}=\dfrac{l_{ij}^2-\gamma_i^2-\gamma_j^2}{2\gamma_i\gamma_j} $.
    \item \textbf{end for}
\end{enumerate}
\end{algorithm}
In the second step, by optimizing Ricci energy $E(u) = \int{\sum_i(\bar{K_i}-K_i)}du_i $, where $u_i = log \gamma_i$ and $u = (u_1,u_2,...,u_n)^T$ (n is the number of vertices), the desired metric is determined for the preset target curvature. Traversing all faces, for each face $f_{ijk}$, calculate the power circle that is orthogonal to three vertex circles and define the power center as $o_{ijk} $ and the distance from $o_{ijk} $ to $e_{ij}$ by $h_{ij}^k$. By traversing all the edges, for each edge $e_{ij}$, if it is adjacent to two faces $f_{ijk}$ and $f_{jil}$, then its weight is given by 
\begin{equation} \label{edgeWeightEqu1}
	w_{ij} = \dfrac{h_{ij}^k+h_{ji}^l}{l_{ij}}.
\end{equation}
If the edge is attached to a single face  $f_{ijk}$, then
\begin{equation} \label{edgeWeightEqu2}
	w_{ij} = \dfrac{h_{ij}^k}{l_{ij}}.
\end{equation}
Then the Ricci energy's Hessian matrix is then constructed as  $H = \dfrac{\partial^2 E}{\partial u_i \partial u_j} $, where 
\begin{equation} \label{HessianEqu}
	\dfrac{\partial^2 E}{\partial u_i \partial u_j} = \begin{cases} 
	\sum_k w_{ik} \quad i = j\\
	-w_{ij} \quad e_{ij} \in M   \quad . \\	
	0 \quad else
	\end{cases}
\end{equation}
And Newton's method is used to optimize the energy $E(u)$ constrained on the hyperplane $\sum_i u_i = 0$,
\begin{equation}
	\delta u = H^{-1}\triangledown E(u) = H^{-1}(\bar{K}-K).
\end{equation} 
The pipeline of this phase is shown in Algorithm \ref{algorithm2}.
\begin{algorithm}[!h] \fontsize{10pt}{10pt}\selectfont
\caption{Ricci energy optimization}\label{algorithm2}
\textbf{Input:} $ M , \bar{K} $: (a triangular mesh , the preset curvature).\\
\textbf{Output:} $ g $: A desired metric for the preset curvature $\bar{K} $.
\begin{enumerate}
\item Using Algorithm \ref{algorithm1}, calculate the initial circle packing metric.
\item \textbf{do}
\item   \quad calculate the power circle center $ o_{ijk} $ for all of $ f_{ijk}$, then calculate the distance to the edges of $ f_{ijk}$ : $ h_{ij}^k, h_{jk}^i $ and $ h_{ki}^j $.
\item   \quad For all of $e_{ij} $, calculate the edge weight $ w_{ij} $ using \ref{edgeWeightEqu1} and \ref{edgeWeightEqu2}.
\item	\quad Create the Hessian matrix with the help of \ref{HessianEqu}.
\item 	\quad Solve linear equation $ H\delta u = \bar{K} - K $ constrained on $ \sum_i u_i = 0  $.
\item   \quad Update conformal factor $ u \leftarrow u + \delta u $.
\item 	\quad For all of $e_{ij} $, calculate the length of the edge by $ l_{ij}^2 = \gamma_i^2 + \gamma_j^2 + 2\gamma_i\gamma_j\eta_{ij} $.
\item   \quad For all of $ f_{ijk}$, calculate the corner angles $ \theta_i^{jk} $, $ \theta_j^{ki} $ and $ \theta_k^{ij} $ by cosine law.
\item   \quad For all $v_i $ of, calculate the Gaussian curvature by $ K_i = 2\pi - \sum_{jk} \theta_i^{jk}$.
\item \textbf{while} $ max_{v_i\in M}|\bar{K_i}-K_i|<\epsilon $
\end{enumerate}
\end{algorithm}

The mesh with the desired metric is finally inserted into the plane. First, an initial face $f_{ijk}$ is chosen randomly and flattened isometrically to the plane,
\begin{equation}\label{embed1}
	\phi(v_0) = 0, \phi(v_1) = l_{01}, \phi(v_2) = l_{20}e^{i\theta_0^{12}}.
\end{equation}
All neighboring faces are then placed in a queue. The head face $f_{ijk}$ is ejected when the queue is not empty. Assume $\phi(v_i)$ and $\phi(v_j)$ have been calculated. Consequently, $\phi(v_k)$ is at the point where the two circles $(\phi(v_i),l_{ik})$ and $(\phi(v_j),l_{jk})$ cross, furthermore, $\phi(v_k)$ is chosen to maintain the flattened face's normal upward. For each face $f_{ijk}$  in the mesh, a queue is created and all of its neighboring faces that have not yet been flattened, are added to the queue. Until the queue is empty, the procedure of flattening the faces is repeated. Algorithm \ref{algorithm3} outlines the steps for completing it.
\begin{algorithm}[!h] \fontsize{10pt}{10pt}\selectfont
\caption{Embedding in the plane}\label{algorithm3}
\textbf{Input:} $ M $: a triangular mesh .\\
\textbf{Output:} $ \phi $: an isometric embedding.
\begin{enumerate}
\item Embed a randomly selected initial triangle $ f_{ijk} \in M $ using \ref{embed1}.
\item Create the queue $Q$ with all the initial face's surrounding faces.
\item \textbf{while} $\mid Q \mid > 0$ \textbf{do}
\item \quad Pop the head face $ f_{ijk} $ from the queue $Q$.
\item \quad Assume that $ v_i $ and $ v_j $ have been embedded. Determine the points where two circles overlap, $ (\phi(v_i),l_{ik}) \cap (\phi(v_j),l_{jk}) $. 
\item \quad Add to the queue the nearby faces of $ f_{ijk}$ that have not yet been accessed.
\item \textbf{end while}
\end{enumerate}
\end{algorithm}
\newpage
\section{Proposed method}\label{sec4}
\paragraph{Overview} 
This study investigates the potential of employing manifold-based classification techniques on covariance matrices obtained through Ricci energy optimization. The goal is to aid in the diagnosis of Alzheimer's disease by analyzing hippocampal regions within brain surface data. The process begins with reconstructing the brain surface from MRI images using the Freesurfer pipeline. Subsequently, the hippocampal region is extracted from this brain surface. To establish a conformal correspondence between a boundary surface and a plane, a planar conformal parametrization is computed via Euclidean Ricci flow. During the Ricci energy optimization process, feature points are uniformly selected. At each point, various geometric characteristics are computed, encompassing the conformal factor, area distortion, and heat kernel. 

Rather than utilizing the individual geometric features themselves, we suggest generating covariance matrices for these features. These covariance matrices contain certain aspects of the local geometry, which may differ in type, dimension, or scale. A notable advantage is that these covariance matrices inherently combine the diverse properties without requiring normalization.

\begin{figure*}[!h]
	\centering
	\includegraphics[width=0.8\textwidth]{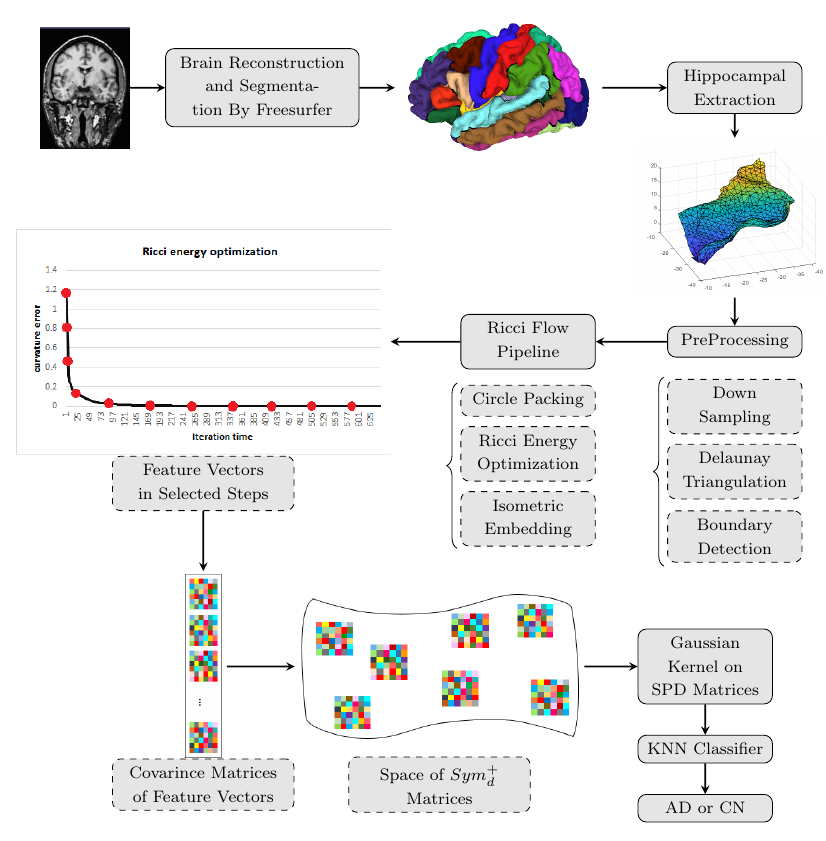}
	\caption{\textcolor{black}{Block diagram of the proposed method in this paper.}}
    \label{blockDiagram}
\end{figure*}

On the other hand, these covariance matrices reside within the Symmetric Positive Definite (SPD) manifold, a space devoid of traditional Euclidean structures such as norm and inner product. Consequently, conventional classification methods in their original forms cannot be employed.
For this, we inspire from recent research that employed kernel approaches on manifold-based data \cite{tabia2015covariance, yun2016exploiting} and examine, for the first time, its application to brain surface categorization. We use a Gaussian radial basis function (RBF) that transforms covariance matrices into an infinite-dimensional Hilbert space and use KNN to classify covariance matrices. In Figure \ref{blockDiagram}, we provide a summary of the suggested approach.
\subsubsection*{Conformal factor}\label{subsubsec9}
\begin{figure}[!h]
    \centering
    \begin{subfigure}{0.7\textwidth}
    \includegraphics[width=\textwidth]{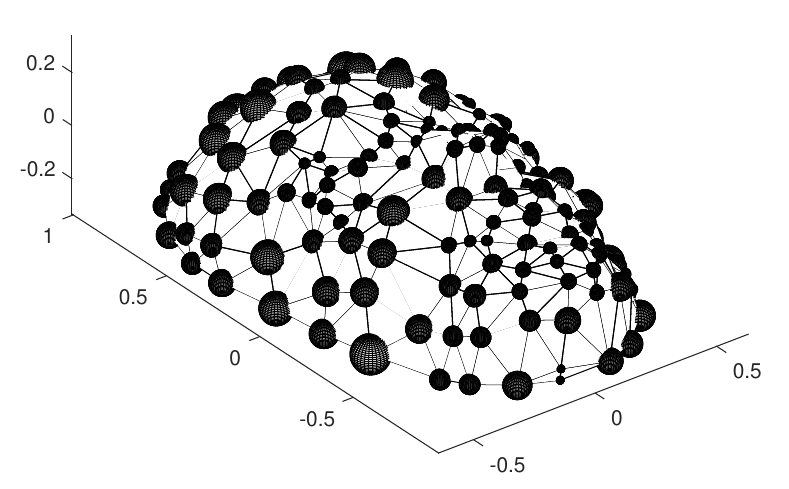}
    \subcaption{\textcolor{black}{A 3D triangular domain covered by an inversive distance circle packing metric before deformation.}}
    \end{subfigure}
    \begin{subfigure}{0.6\textwidth}
    \includegraphics[width=\textwidth]{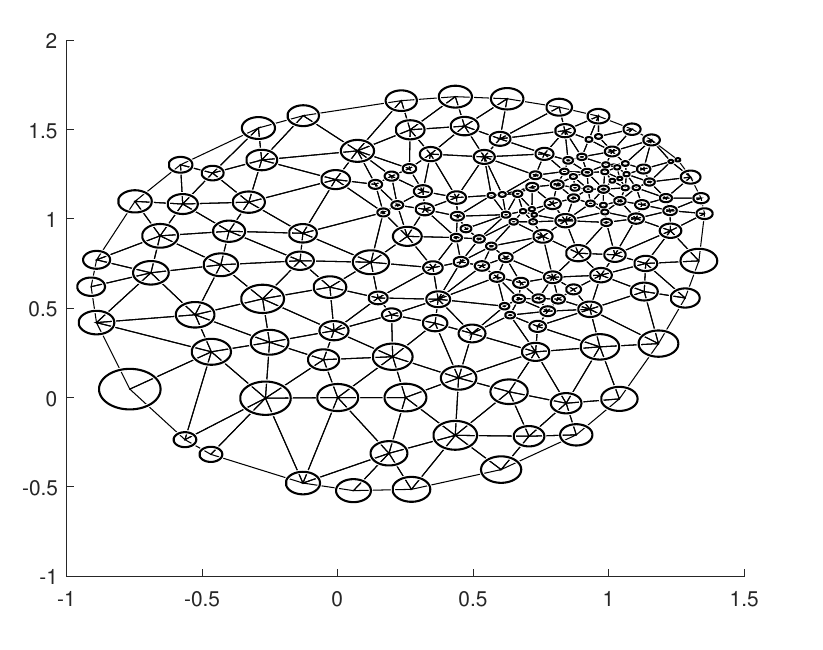}
     \subcaption{A 2D suitable circular range after deformation by Ricci flow.}
    \end{subfigure}
    \caption{Discrete conformal mapping using circle packing.}
    \label{MaskEmbedding}
\end{figure}

The Ricci flow method achieves a conformal parameterization by mapping a 2-manifold to a Euclidean plane, a 2-sphere, or a hyperbolic plane (depending on the manifold topology) with constant Gaussian curvature (0, +1, or -1, respectively). This convergence is illustrated in Figure \ref{MaskEmbedding}. Initially, circle packing metrics determine Gaussian curvature, which aligns with the target curvature via Ricci flow. The process adjusts circle radii, altering Gaussian curvature until it is uniform across vertices. 
In this process, the conformal factor at the vertex $v_i$, is defined as $u_i=\log(\gamma_i)$ where $\gamma_i$ denotes the circle's radius at that vertex. By definition, $u_i$ encodes intrinsic surface information and is invariant to isometries. 
\subsubsection*{Area distortion}\label{subsubsec10}
\begin{figure}
    \centering
    \begin{subfigure}{0.48\textwidth}
    \includegraphics[width=\textwidth]{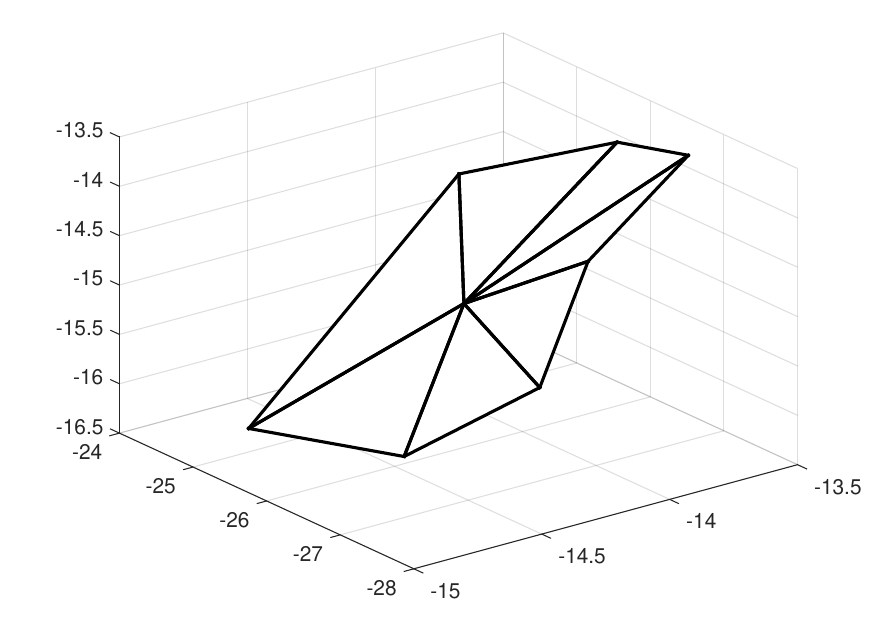}
     \caption{}
    \end{subfigure}
    \begin{subfigure}{0.45\textwidth}
    \includegraphics[width=\textwidth]{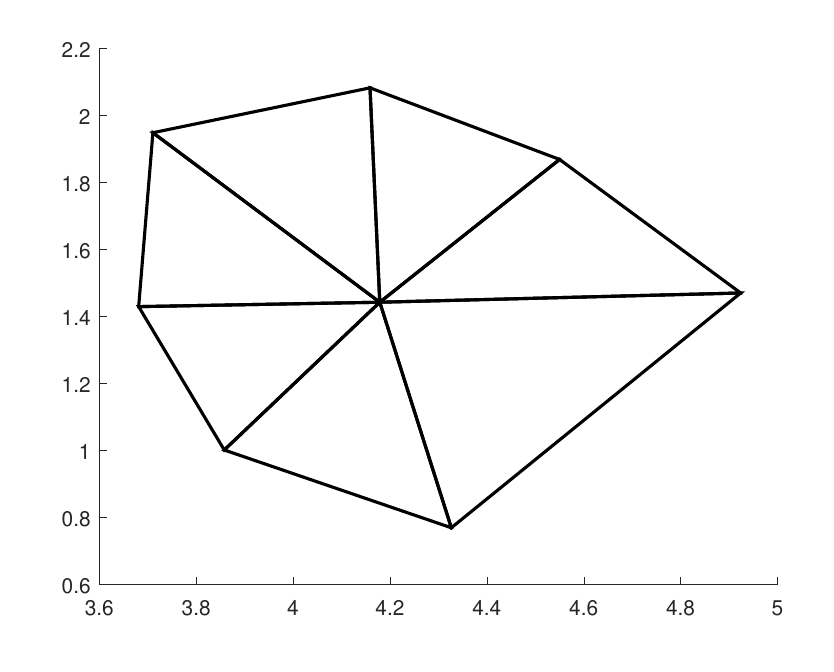}
     \caption{}
    \end{subfigure}
    \caption{(a): The local area around a vertex in the first stage (3-dimensional), (b): The local area around a vertex in the last stage (2-dimensional) of Ricci energy optimization.}
    \label{localArea}
\end{figure}

As explained earlier, the Riemannian metric is deformed by the Ricci flow in accordance with the curvature, evolving as a heat diffusion process and becoming constant over the surface. Therefore, different curvatures on vertices make different metrics on edges and we can use metric as a feature, and compute the area of triangles in a the one-ring local neighborhood at each vertex on a mesh in the first and current stages of the optimization. Then, we calculate the difference of local areas in these two stages and call it the \textit{area distortion} as equation \ref{areaDistortion}.
\begin{equation} \label{areaDistortion}
AD(v) = \sum_{t \in B} area(t)-\sum_{t \in B} \widehat{area}(t),
\end{equation}
where $ area(t) $ is the area of the triangle $ t $ on the initial mesh, $ \widehat{area} $ is the
area of the triangle $ t $ on mesh in the current stage, and $ B $ is one-ring neighborhood of vertex $ v $.  Figure \ref{localArea} shows the local one-ring neighborhood of a vertex in both the first and last stages of Ricci flow optimization.
\subsubsection*{Heat kernel}
Because the heat kernel represents the entire invariant of the Riemannian metric, it plays a crucial role in geometric modeling and shape analysis \cite{sun2009concise}.
The heat equation governs the heat diffusion process on $M$. The temperature field on $M$ at time $t$, represented as $u(x,t):M\times\mathbb{R}^{+}\longrightarrow\mathbb{R}$, satisfies the heat equation given below:
\begin{equation} \label{hkequaton1}
\Delta_{g}u(x,t) =- \frac{\partial u(x,t)}{\partial t},
\end{equation}
with the initial condition $u(x,0)$.\\
The heat kernel can explicitly provide the answer to the heat equation \eqref{hkequaton1}:
\begin{equation}
u(x,t)=\int_M K(x,y,t)u(y,0)dy.
\end{equation}
Since in this work we focus on discrete surfaces, we compute a discretized version of the heat kernel using the discretized Laplace-Beltrami operator. Cotangent edge weight is used to build the discrete Laplace-Beltrami operator.\\
\textbf{Definition (cotangent edge weight):} The well-known cotangent edge weight \cite{dodziuk1976finite,pinkall1993computing} on a triangle mesh is defined as follows:
\begin{equation}
w_{ij} = \begin{cases}
\frac{1}{2}\cot \alpha , \,\,\quad\quad\quad\quad [v_i,v_j]\in {\partial M} \\
\frac{1}{2}(\cot \alpha + \cot \beta) , \quad [v_i,v_j]\notin {\partial M}  .
\end{cases}
\end{equation}
\textbf{Definition (discrete Laplace matrix):} The discrete Laplace matrix $L = (L_{ij})$ for a Euclidean triangle mesh is given by
\begin{equation}
L_{ij} = \begin{cases}
-w_{ij}, \quad\quad i \neq j \\
\sum_{k}w_{ik}, \quad i = j .
\end{cases}
\end{equation}
Because $ L $ is symmetric, it can be decomposed as 
\begin{equation}
L = \Phi\Lambda\Phi^{T},
\end{equation}
where $ \Lambda = \text{diag}\left(\lambda_{0}, \lambda_{1},..., \lambda _{n}\right) , 0=\lambda_{0}<\lambda_{1}\leqslant\lambda_{2}\leqslant ... \leqslant \lambda_{n} $, are the eigenvalues of $ L $, and $ \Phi = (\phi_{0}-\phi_{1}-\phi_{2}-...-\phi_{n}) $, $ L\phi_{i} = \lambda_{i}\phi_{i} $ are the orthonormal eigenvectors, $ n $ is the number of vertices, such that $ \phi_{i}^{T}\phi_{j}=\delta_{ij}$.\\
Consequently, the discrete heat kernel is described as follows:
\begin{equation}
HK(t) = \Phi exp(-\Lambda t) \Phi^{T}.
\end{equation}
There are several good qualities to the heat kernel function \cite{sun2009concise}. The intrinsic property, the informative property, and the multi-scale property are the key properties that we are interested in. The intrinsic property states that the heat kernel is invariant under isometric transformation, allowing the Laplacian to be represented in local coordinates as a function of metric. According to the informative quality, the heat kernel has complete knowledge of the intrinsic geometry of a Riemannian manifold $ M $ and can fully describe its shape up to isometry.\\
As a result, depending on the scale parameter $ t $ used, the heat kernel can describe the local shape and is an interesting candidate for a point signature. However, one major drawback of utilizing the family of functions $ {HK_{t}(x,.)}_{t>0} $ to describe point $ x $ is its high computational complexity.

\begin{figure*}[!h]
    \centering
    \begin{subfigure}{0.49\textwidth}
    \includegraphics[width=\textwidth]{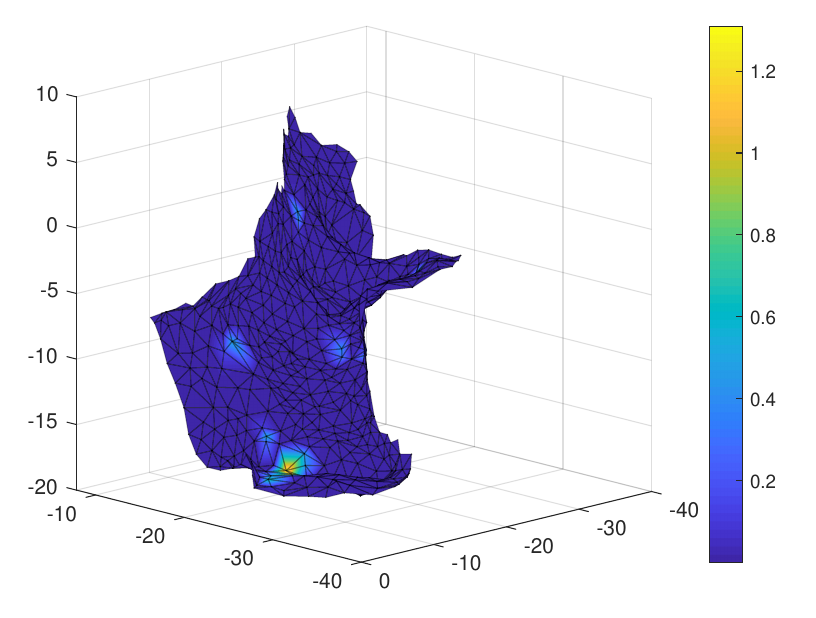}
    \end{subfigure}
    \begin{subfigure}{0.49\textwidth}
    \includegraphics[width=\textwidth]{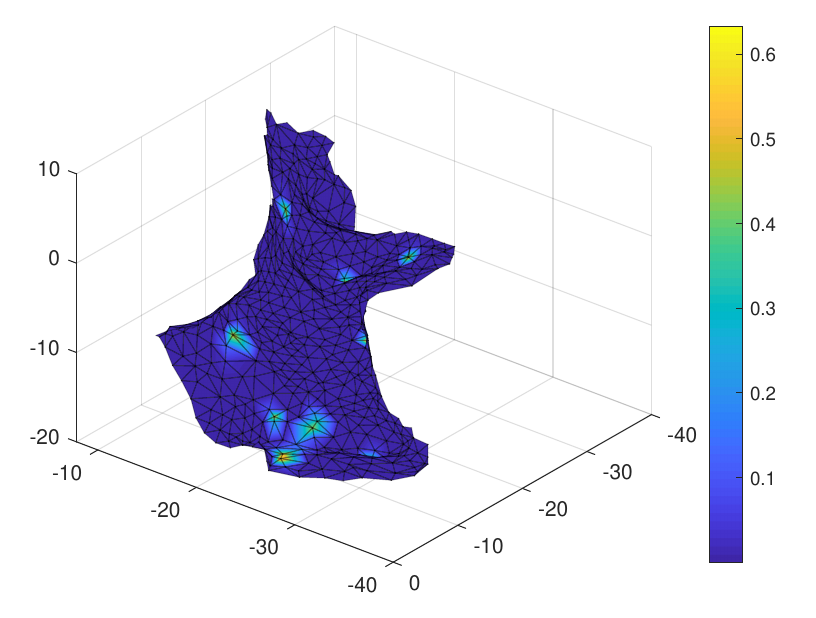}
    \end{subfigure}  
    \caption{\textcolor{black}{Heat kernel distribution on a hippocampal region of the cerebral cortex in t=1 (left) and t=10 (right).}}
    \label{HeatKernelCurv}    
\end{figure*}

Heat kernel on mesh $M$, $ {HK_{t}(x,.)}_{t>0} $, is defined by the product of the temporal and spatial domains $ \mathbb{R}^{+}\times M $. Consequently, the heat kernel of all points on $ M $ require $ \mathbb{R}^{+}\times M \times M$ space, in addition to the expense of matching the neighbours when comparing the heat kernels of two points.\\
There is much redundant information in the heat kernel encoding, and changes in the spatial domain of the heat kernel's function are reflected in changes to its temporal behavior. Restricting the heat kernel to the temporal domain and just its subset is one method for overcoming the challenge above. The function $ HKS(x) $ is defined as a function over the temporal domain for a point $ x $ on the manifold $ M $:
\begin{equation}
HKS(x) : \mathbb{R}^{+}\longrightarrow\mathbb{R}, HKS(x,t) = HK_{t}(x,x).
\end{equation}
As demonstrated in \cite{sun2009concise}, under reasonable assumptions, $ {HK_{t}(x,x)}_{t>0} $ retains all of the information of $ {HK_{t}(x,.)}_{t>0} $, even when the signature is limited to the temporal domain and the full spatial domain is dropped.\\
The local curvature of the region surrounding the point $x$ and the $ HKS(x,t) $ specified are closely related. The rate of heat diffusion provides an understandable explanation for the consistency between the Gaussian curvature and the value of the heat kernel function. On a Riemannian manifold, the scalar curvature (twice the Gaussian curvature) measures the difference between the volume of a geodesic ball ($ B_g $) and that of a Euclidean ball ($ B_{\mathbb{R}^n} $) of equal radius. As a result, regions with lower scalar curvature have bigger $B_g$. Heat tends to disperse faster in regions with slight curvature than regions with considerable curvature. Accordingly, with a small fixed $ t $, points locally retain more heat in large curvature regions, leading to higher heat kernel function values.  Figure \ref{HeatKernelCurv} depicts the distribution of heat kernel on a hippocampus region at t=1 and t=10.
\subsection{Riemannian geometry of SPD matrices} \label{subsec3}
This section discusses the geodesic distance, which has been used as a kernel metric on the Riemannian manifold of SPD matrices, as well as the mathematical properties of the space of covariance matrices.
\subsubsection{The space of covariance matrices} \label{subsubsec7}
The covariance matrix is a matrix used to represent the covariance between different variables. It measures how much two random variables vary together, and is typically used in regression analysis. It can also be used to measure the linear relationship between two variables. Let $M =(Sym_d^+)$ be the set of all non-singular covariance matrices that are SPD matrices with the size $d\times d$. A non-linear Riemannian manifold is a $(Sym_d^+)$ and has an inner product $<.,.>_{X \in M}$ that smoothly varies from point to point for each tangent space $T_X M$ at each point $X \in M$. For the tangent vectors $\Omega \in T_X M$, the inner product creates a norm such that $||\Omega||^2 = <\Omega,\Omega>_X$. A geodesic is the shortest curve that connects two points $X$ and $Y$ on the manifold.
A suitable metric for quantifying the dissimilarity of the covariance matrices $X$ and $Y$ is the length $d_{g}(X, Y)$ of the geodesic between $X$ and $Y$. For every $\Omega \in T_X M$, there exists a distinct geodesic beginning at $X$ in the tangent vector's direction $\Omega$.
The exponential map $exp_X:T_X M \to M $ maps elements $\Omega$ on the tangent space $T_X M$ to the points
$Y \in M$. The length of the geodesic connecting $X$ to $Y$ is given by $d_g(X,exp_X(\Omega))=||\Omega||_X$, \cite{tuzel2008pedestrian}.
\subsubsection{Distance between SPD matrices}
The Riemannian metric of the tangent space $T_X$ at a point $X$ is given as 
$<\Omega,z>_X = trace(X^{\frac{-1}{2}}\Omega X^{-1} z X^{\frac{-1}{2}}) $ \cite{tuzel2008pedestrian}. The exponential map associated with the Riemannian metric $exp_X(\Omega)=X^{\frac{1}{2}}exp(X^{\frac{-1}{2}}\Omega X^{\frac{-1}{2}})X^{\frac{1}{2}}$ is a global diffeomorphism (a one-to-one, onto, and continuously differentiable mapping in both directions). Thus, its inverse is uniquely defined at every point on the manifold: $log_X(Y)=X^{\frac{1}{2}}log(X^{\frac{-1}{2}} Y X^{\frac{-1}{2}})X^{\frac{1}{2}}$. Ordinary matrix exponential and logarithm operators are denoted by the symbols $ exp $ and $ log $, whereas manifold-specific operators $ exp_X$ and $ log_X$ rely on the point $X \in (Sym_d^+)$. The tangent space of $(Sym_d^+)$ is the space of $d\times d$ symmetric matrices, and the manifold and tangent spaces have the same size $m=d(d+1)/2$.
The standard matrix exponential and logarithm operators can be calculated in the manner shown below for symmetric matrices. Let $X=UDU^T$ represent the eigenvalue decomposition of the symmetric matrix $X$. The definition of the exponential series is: $exp(X)=\sum_{k=0}^{\infty}\frac{X^k}{k!} = U exp(D)U^T$, where the exponential eigenvalue diagonal matrix is $exp(D)$. In a similar manner, $log(X)=\sum_{k=1}^{\infty}\frac{-1^{k-1}}{k}(X-I)^{k} = U log(D)U^T$ provides the logarithm.
The exponential operator is always defined, but the logarithms only exist for symmetric matrices with strictly positive eigenvalues. Then, we can calculate the geodesic distance between any two points on $Sym_d^+$ by: \cite{tuzel2008pedestrian}
\begin{equation} \label{geoDis}
d_{g}^{2}(X,Y) = <log_X(Y), log_X(Y)>_{X} = trace(log^2(X^{-\frac{1}{2}}YX^{-\frac{1}{2}})).
\end{equation}
\subsection{Covariance matrices on Ricci energy optimization}\label{subsec4}
Suppose $ P = \{p_i, i=1,2,...,m\} $ be the set of the selected points (stages) in the Ricci energy optimization on a surface. We compute the \textit{conformal factor}, the \textit{area distortion} $(AD_i)$ and the \textit{heat kernel} $(HK_i)$ signatures in each point $ p_i $ as $ F_i = \{u_i, AD_i, HK_i\} $, which it represents how the Ricci energy optimization changes local geometric properties. We characterize each surface by a set of the covariance matrices $C_i := \dfrac{1}{n} \sum_{j=1}^{n}(F_j-\mu)(F_j-\mu)^T$, $ i=1,2,...,m $, where $ \mu $ is the mean of the feature vectors $ \{F_j\}_{j=1...n} $ computed in the point $ p_i $, and $ n $ is the number of the selected vertices on triangle mesh of the surface. Each feature's variance is represented by the diagonal entries of $ C_i $, whereas each co-variation is represented by the non-diagonal entries.
\subsection{3D shape classification}
\label{3Dclassify}
Once covariance matrices have been computed from feature vectors and the geodesic distance has been defined, the brain surface classification task is covariance classification. However, due to $(Sym_d^+)$ non-linear nature, it is impossible to classify covariance matrices using standard procedures such as K-Nearest Neighbor (KNN). We address this problem using a Gaussian radial basis function (RBF) that transforms the covariance matrices into an infinite-dimensional Hilbert space. Intuitively, this produces a pretty detailed depiction. The Gaussian kernel is represented as follows in $ \mathbb{R}^{d}$:
\begin{equation} \label{GaussKernel}
\kappa(x_i, x_j) = exp(\frac{\parallel x_{i}-x_{j} \parallel ^2 }{2\sigma ^2}),
\end{equation}
which utilizes the Euclidean distance between the two points $x_i$ and $x_j$. By replacing a geodesic distance instead of the manifold's Euclidean distance, we aim to define a kernel on a Riemannian manifold. The benefit of computing positive definite kernels on the SPD matrices' Riemannian manifold is that we may immediately apply algorithms created for $ \mathbb{R}^d$ while still taking into account the geometry of the manifold.
Here, we employ $ \kappa(. , .) $ as the kernel function, $ \mathbb{H} $ as the Hilbert space, and $ \Phi $ as the function of the mapping matrix $C$ in $ \mathbb{H} $. The function $ \Phi $ is not explicitly stated in this study, and by changing the Euclidean distance by the Riemannian manifold's distance provided by Eq. \eqref{geoDis}, the Gaussian kernel is applied as follows:
\begin{equation}\label{MatGaussKernel}
\kappa(C_i, C_j) = exp(\frac{d_{g}^{2}(C_i, C_j)}{2\sigma ^2}),
\end{equation}
where $\Phi$ is a mapping from $M$ to $\mathbb{H}$ such that $ d_{g}^{2}(C_i, C_j) = \parallel \Phi(C_i)- \Phi(C_j) \parallel _{H} $.
\subsection{Surface classification using K-nearest neighbor on Riemannian manifold}
For computing similarity and classification, we build a global distance function so that
one can compare two embedded surfaces by using the covariance descriptors. Given two surfaces represented by two sets of covariance matrices $ S_1  = \{C_i\}_{i=1,...,m}$ and $ S_2 = \{C_j\}_{j=1,...,m}$, which $ m $ is the number of the selected stages from Ricci energy optimization. First, a matrix of similarity scores between $S_1$ and $S_2$ is computed. The RBF-kernel provided by Eq. \eqref{MatGaussKernel} is a popular option for the similarity measure, sometimes known as the minor kernel. The kernel value can then be determined by averaging the $S_1$ and $S_2$ components with the highest matching scores as follows:
\begin{equation} \label{distanceK}
K(S_1, S_2)=\frac{1}{2}[\widehat{K}(S_1,S_2)+\widehat{K}(S_2, S_1)],
\end{equation}
where $ \widehat{K}(S_1, S_2) = \frac{1}{|S_1|}\sum_{i=1}^{|S_1|} min_{j=1,...,|S_2|} \kappa(C_i, C_j)$.\\
Then we propose to use K-nearest neighbor (KNN) using this similarity measure to classify a surface. An algorithm of the proposed method can be seen in Algorithm \ref{algorithm}.
\begin{algorithm}[!h] \fontsize{10pt}{10pt}\selectfont
\caption{The proposed method algorithm}\label{algorithm}
\textbf{Input:} A triangular mesh data $M \in \mathbb{R}^3$.\\
\textbf{Output:} The class number of input mesh data $M$.
\begin{enumerate}
    \item Calculate the inversive distance circle packing metric of mesh $M$ using Algorithm \ref{algorithm1}.
    \item Optimize the Ricci energy of $M$ by Newton's method using Algorithm \ref{algorithm2} and calculate the conformal factor, area distortion, and heat kernel signatures on the vertex $v_i$ from the selected vertices of mesh $M$ in the selected steps in Ricci energy optimization as $ u_i, AD_i, HK_i $.
    \item Calculate covariance matrix of features in each step $i$ of Ricci energy optimization on selected vertices as $C_i$.
    \item Move the covariance matrices to an infinite-dimensional Hilbert space using Gaussian radial basis function (RBF) in Equation \ref{MatGaussKernel}.
    \item Classify mesh data $M$ by KNN classifier using distance function in Equation \ref{distanceK} based on the covariance matrices in the Hilbert space.
\end{enumerate}
\end{algorithm}

\section{Experimental setting and results}
By examining the brain cortices of people with Alzheimer's disease (AD) and cognitively normal (CN) individuals, we show the efficacy and efficiency of our method. Triangular meshes are used to depict the brain's cortical surfaces. 
The implementation of our proposed method was carried out using MATLAB on a Windows 10 platform on a computer equipped with a 3.00 GHz Intel(R) Core(TM) i7-9700 processor and 16 GB RAM. In the following sections, we will discuss the data source, preprocessing, results, and performance comparison.
\subsection{Data source}
The data used in this paper's study was sourced from the Alzheimer's Disease Neuroimaging Initiative (ADNI) database \footnote{https://adni.loni.usc.edu/data-samples/access-data/}. \textcolor{black}{The data were randomly obtained from the ADNI database \cite{mueller2005alzheimer} containing 100 AD and 100 CN subjects, T1-weighted structural MRI scans at initial or baseline screening, including ADNI 1, 2/GO, and 3.} 
\subsection{Data preprocessing}
\begin{figure*}[h!]
\centering
\begin{subfigure}{0.49\textwidth}
\includegraphics[width=\textwidth]{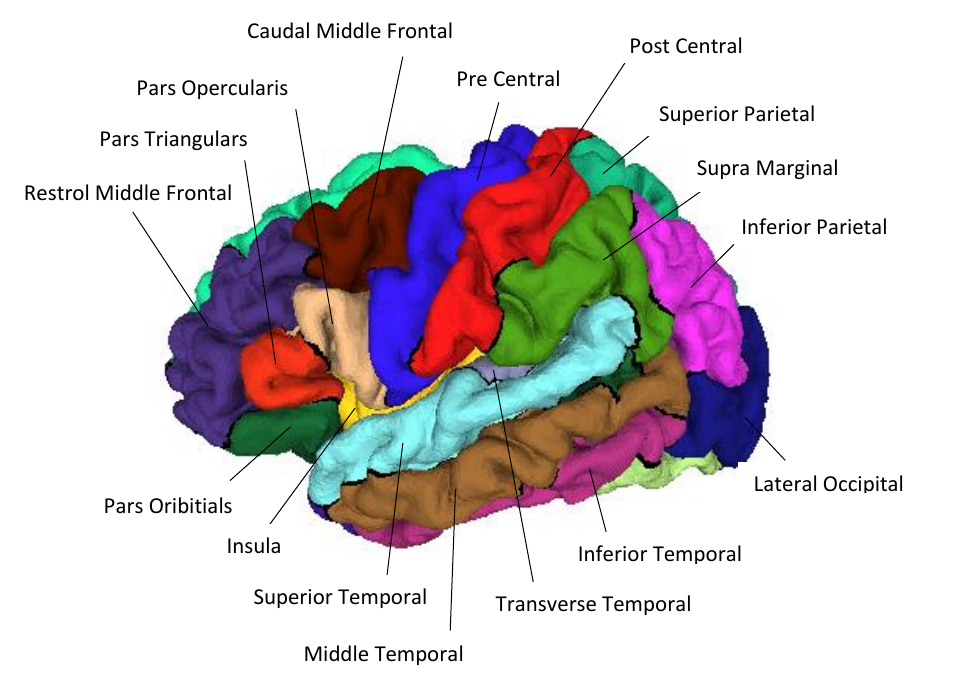}
\caption{superior view}
\end{subfigure}
\begin{subfigure}{0.49\textwidth}
\includegraphics[width=\textwidth]{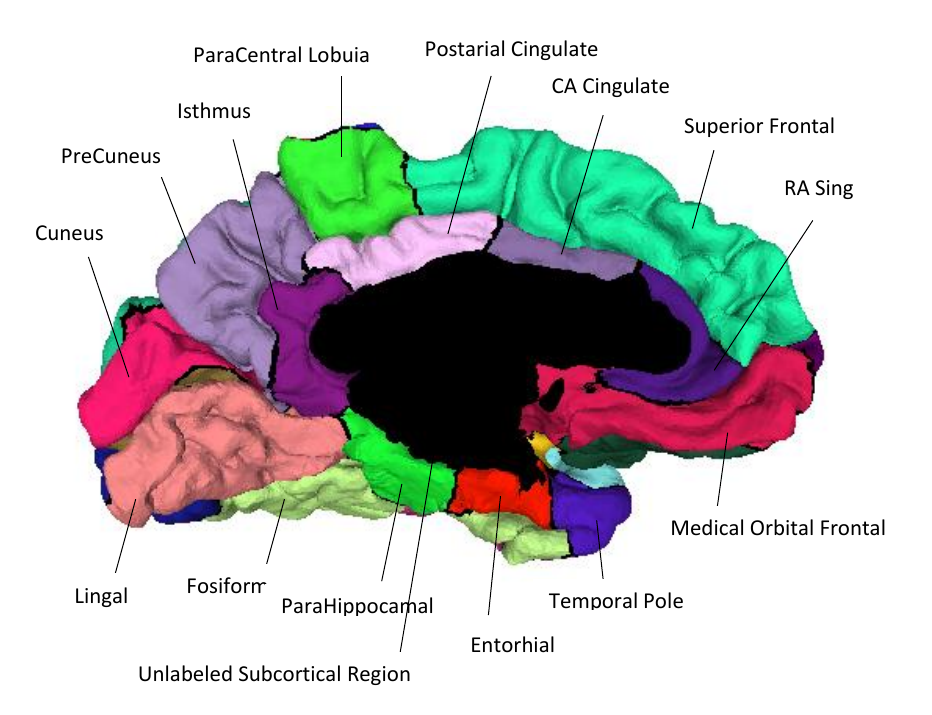}
\caption{Inferior view}
\end{subfigure}
\caption{\textcolor{black}{Illustration of the functional areas of the left hemisphere of the cerebral cortex.}}
\label{brainColoring}
\end{figure*}

We automated the cortical and subcortical parcellations, surface extraction, tissue categorization, and skull stripping using the Freesurfer processing pipeline \cite{dale1999cortical}. It calculates geometric properties, including curvature, curving, and local folding for each parcellation, and offers surface and volume data for around 34 different cortical structures \cite{desikan2006automated}. The left hemisphere of the brain is depicted in Figure \ref{brainColoring} with several functional regions. 

\begin{figure}[!h]
    \centering
    \begin{subfigure}{0.49\textwidth}
    \includegraphics[width = 7cm,height = 4.5cm]{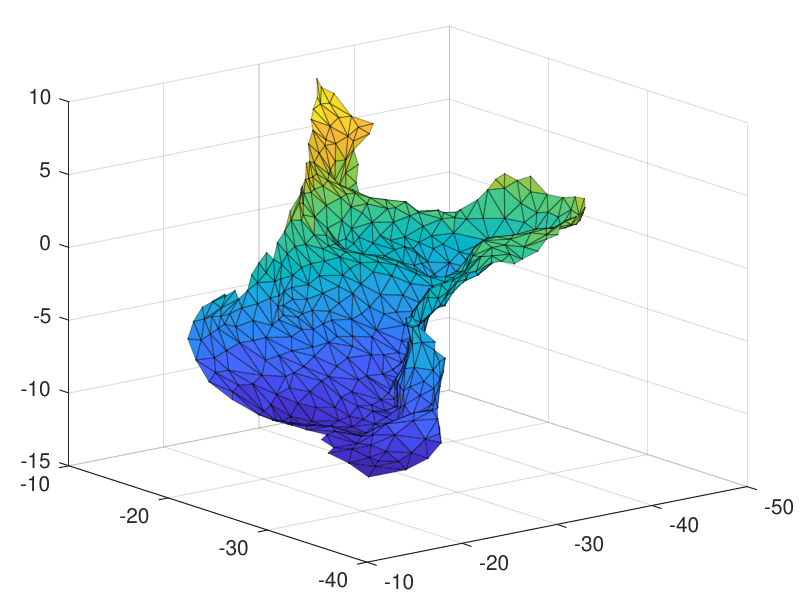}
    \caption{}    
    \end{subfigure}
    \begin{subfigure}{0.49\textwidth}
    \includegraphics[width = 7cm,height = 4.5cm]{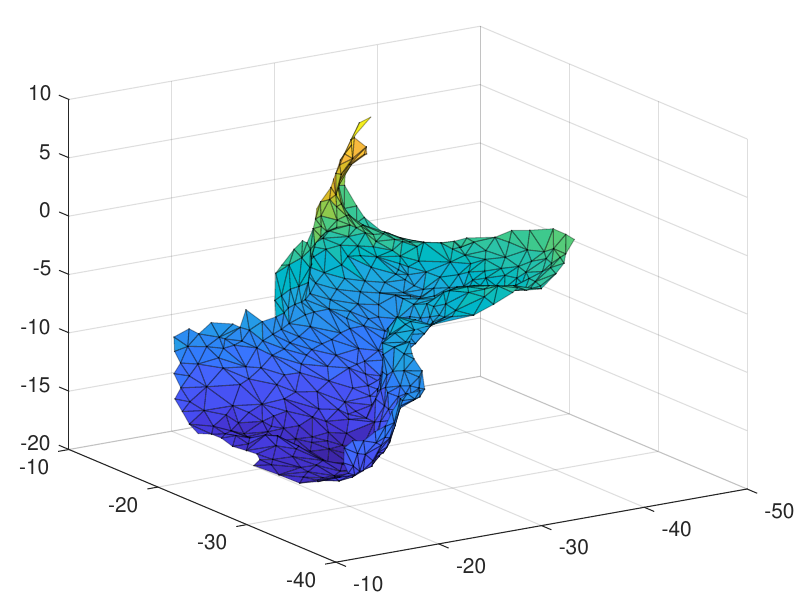}
    \caption{}    
    \end{subfigure} 
    \caption{\textcolor{black}{The hippocampal region of the left hemisphere of the cerebral cortex of an AD (left) and a CN (right) subject.}}
    \label{fig:hippocampalRegionsADCN}    
\end{figure}

According to \cite{desikan2006automated}, we extracted the hippocampus region after labeling several cortical surface functional regions in various colors. Because the hippocampal region of the left hemisphere is more affected by Alzheimer's disease than other regions, we focused on it in this study. The hippocampal regions of the left hemispheres of AD and CN subjects can be seen in Figure \ref{fig:hippocampalRegionsADCN}. 
\subsection{Results}
\begin{figure}[!h]      
    \centering
    \begin{subfigure}{0.49\textwidth}
    \includegraphics[width = 7cm,height = 4.5cm]{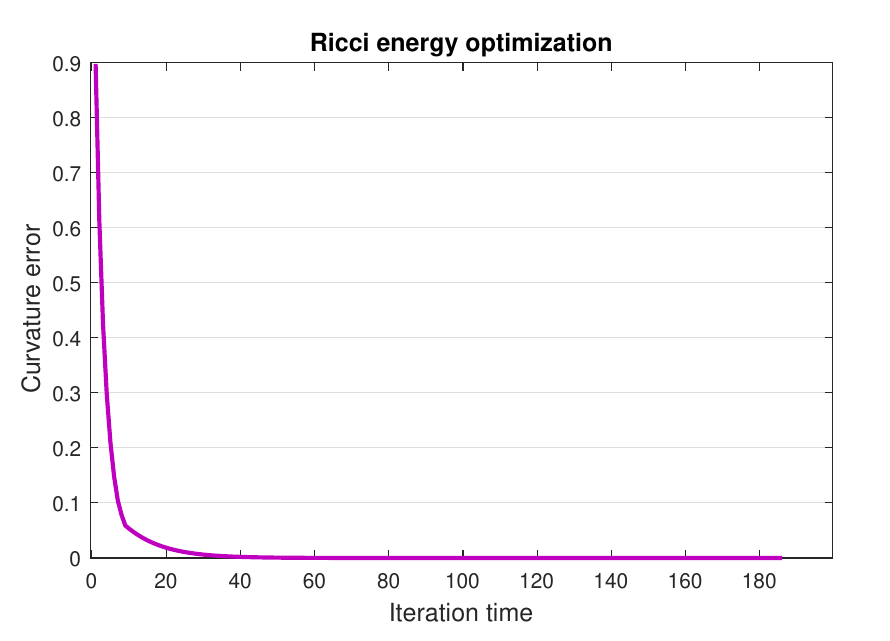}
    \caption{}
    \end{subfigure}  
    \begin{subfigure}{0.49\textwidth}
    \includegraphics[width = 7cm,height = 4.5cm]{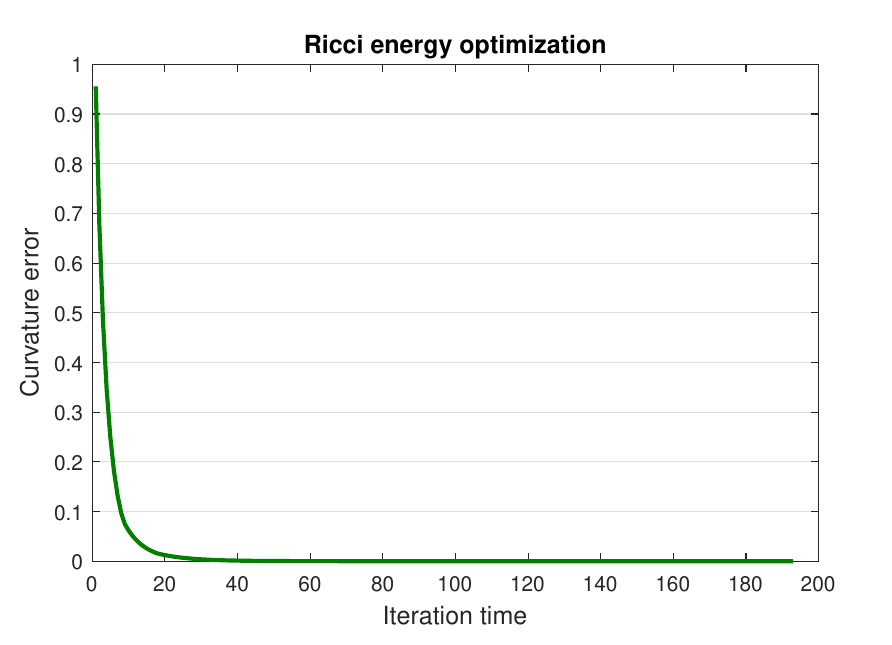}
    \caption{}
    \end{subfigure} 
    \caption{\textcolor{black}{Optimization of Ricci energy on the left hippocampal region of an AD subject (left) and a CN subject (right).}}   
    \label{fig:RicciEnergyOptimizationADCN}    
\end{figure}

\begin{figure}[!h]           
    \centering  
    \begin{subfigure}{0.48\textwidth}
    \includegraphics[width = 7cm,height = 4.5cm]{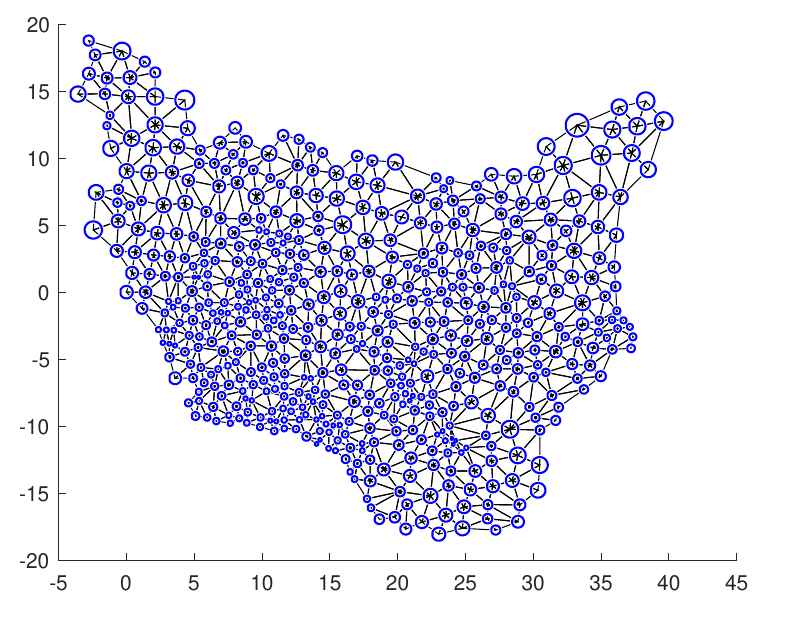}
    \caption{}
    \end{subfigure} 
    \begin{subfigure}{0.48\textwidth}
    \includegraphics[width = 7cm,height = 4.5cm]{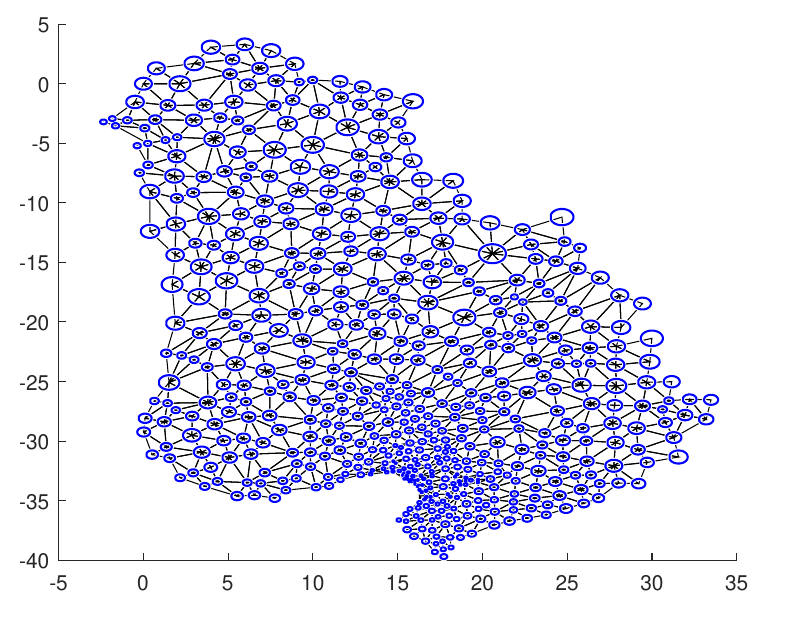}
    \caption{}
    \end{subfigure}   
    \caption{\textcolor{black}{Planar embedding of the hippocampal region of an AD subject (left) and a CN subject (right) after Ricci energy optimization. The blue circles show the inversive distance circle packing in the normalization space.}}
    \label{fig:embeddingADCN}    
\end{figure}

We examined the discrimination power of the suggested shape descriptor on the hippocampal region of a set of left brain hemispheres of 100 CN subjects and 100 AD patients. 
We applied Ricci flow on each region and calculated conformal factor, area distortion and heat kernel features in $ n $ selected stages of the Ricci energy optimization, and computed covariance matrix of these feature vectors and indexed each subject with a set of the covariance matrices. The results of the Ricci flow process and embedding can be seen in Figure \ref{fig:RicciEnergyOptimizationADCN} and Figure \ref{fig:embeddingADCN}, respectively. 

\begin{figure*}[!h]
    \centering
    \begin{subfigure}{0.49\textwidth}
    \includegraphics[width = 7cm,height = 4.5cm]{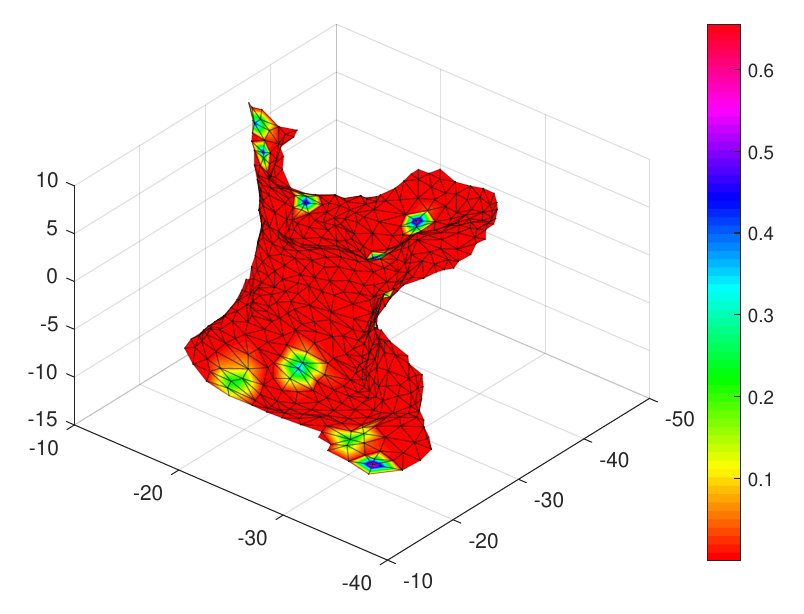}
    \end{subfigure}
    \begin{subfigure}{0.49\textwidth}
    \includegraphics[width = 7cm,height = 4.5cm]{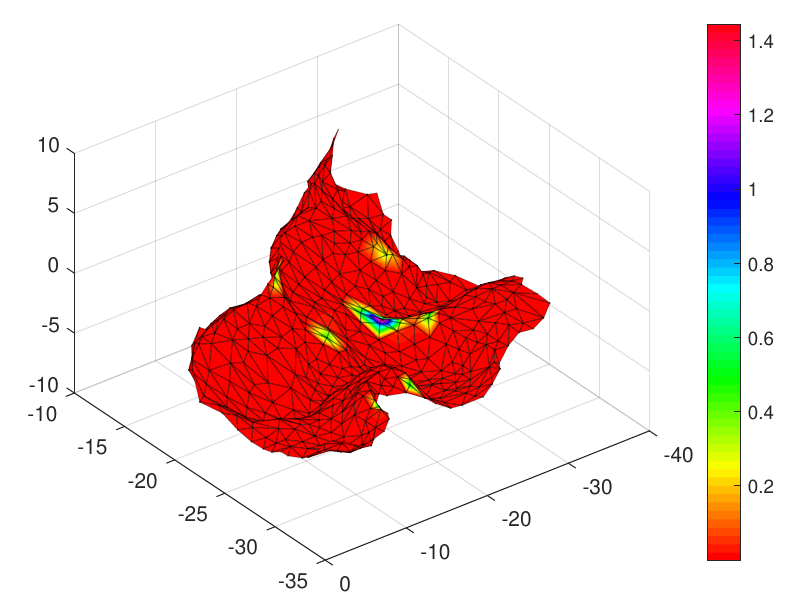}
    \end{subfigure}  
    \caption{\textcolor{black}{Heat kernel distribution on left hippocampal region of AD(left) and CN(right) subjects.}}
    \label{HeatKernel}    
\end{figure*}

\begin{figure*}[!h]
    \centering
    \begin{subfigure}{0.49\textwidth}
    \includegraphics[width = 7cm,height = 4.5cm]{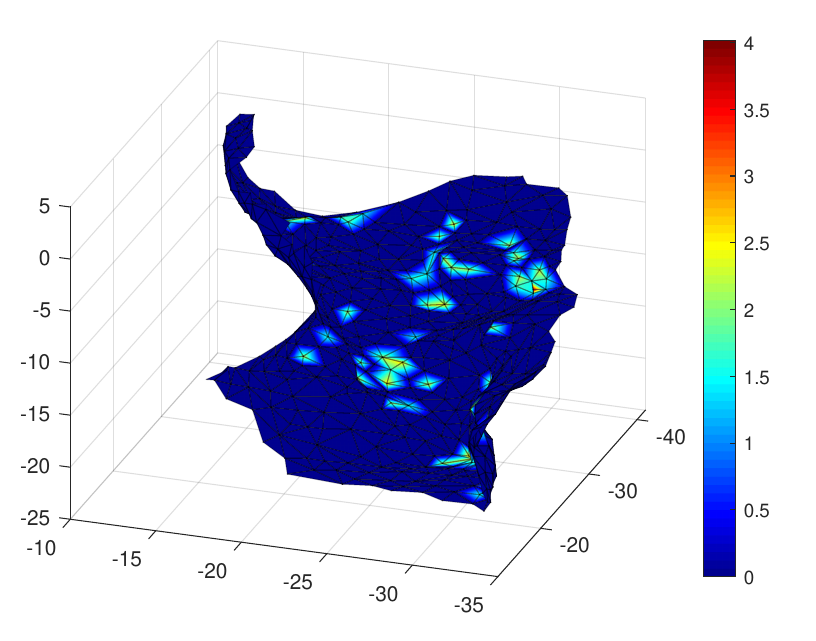}
    \end{subfigure}
    \begin{subfigure}{0.49\textwidth}
    \includegraphics[width = 7cm,height = 4.5cm]{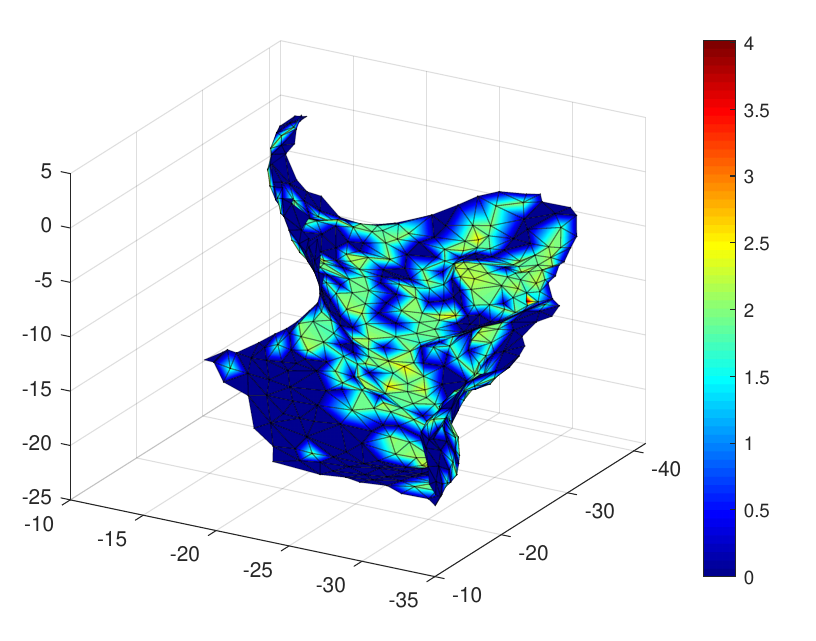}
    \end{subfigure}  
    \caption{\textcolor{black}{Curvature distribution on a left hippocampal region of an AD subject with the different thresholds. Left: Curvatures greater than 0.2 and less than -0.2 highlighted, Right: Curvatures greater than 0.05 and less than -0.05 highlighted.}}
    \label{KThreshOnSurf}    
\end{figure*}

Given a small fixed t,	Figure \ref{HeatKernel} shows the values of heat kernel function on hippocampal region. The	function values are mapped from red (lowest) to crimson (highest).
As shown in Figure \ref{HeatKernel}, big Gaussian curvatures often have large values of the heat kernel function compared to regions with small (negative) curvatures. 
For better correspondence between regions, we considered feature vectors on some vertices according to the curvature using a curvature threshold. Curvature distribution on a left hippocampal region with a different threshold can be seen in Figure \ref{KThreshOnSurf}.

\begin{figure*}[!h]
\centering  
    \begin{subfigure}{0.49\textwidth}
    \includegraphics[width = 7.5cm,height = 4cm]{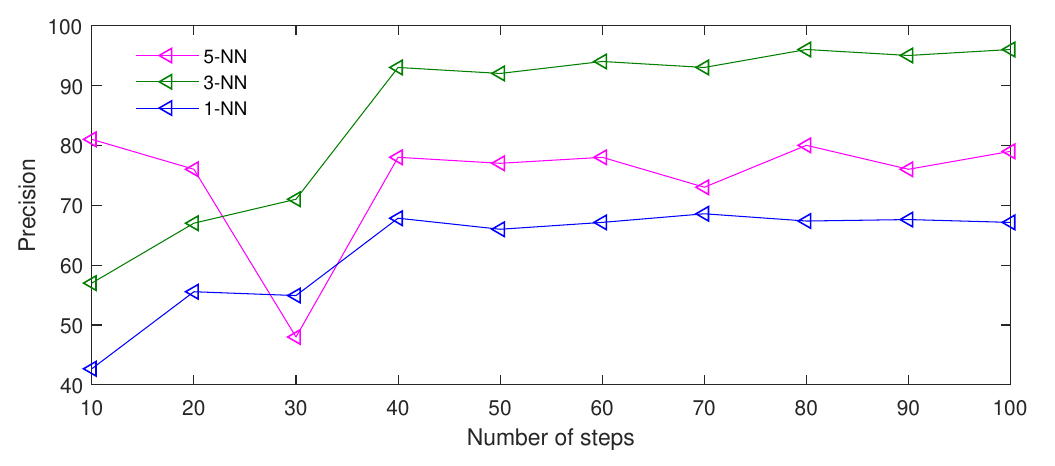}
    \caption{}
    \end{subfigure} 
    \begin{subfigure}{0.49\textwidth}
    \includegraphics[width = 7.5cm,height = 4cm]{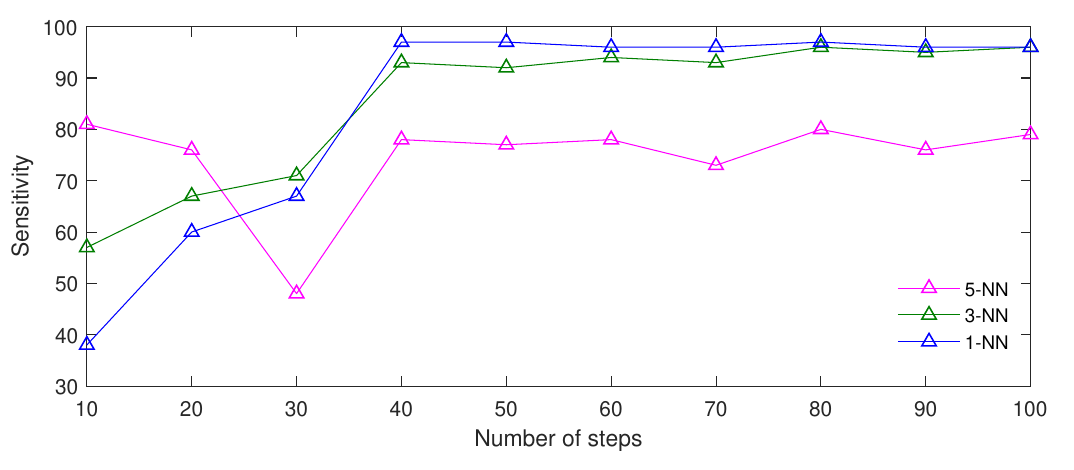}
    \caption{}
    \end{subfigure}  
    \begin{subfigure}{0.49\textwidth}
    \includegraphics[width = 7.5cm, height = 4cm]{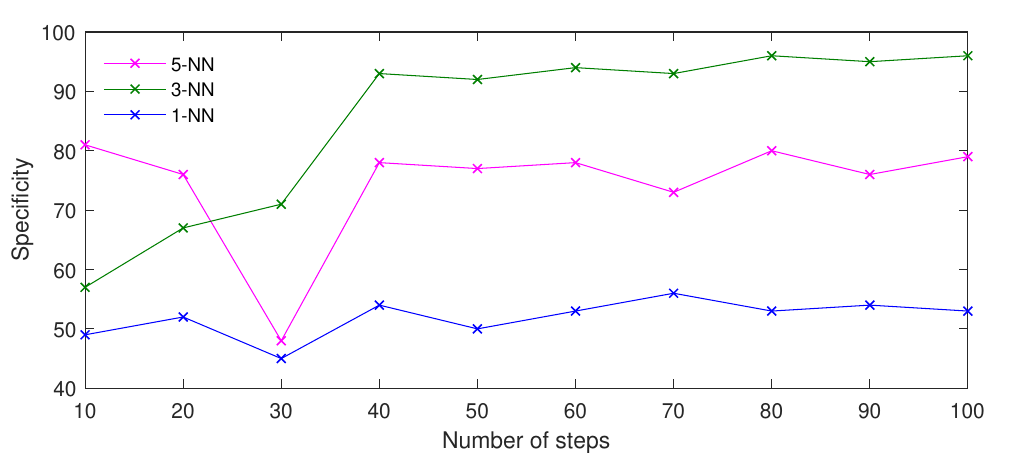}
    \caption{}
    \end{subfigure}  
    \begin{subfigure}{0.49\textwidth}
    \includegraphics[width = 7.5cm, height = 4cm]{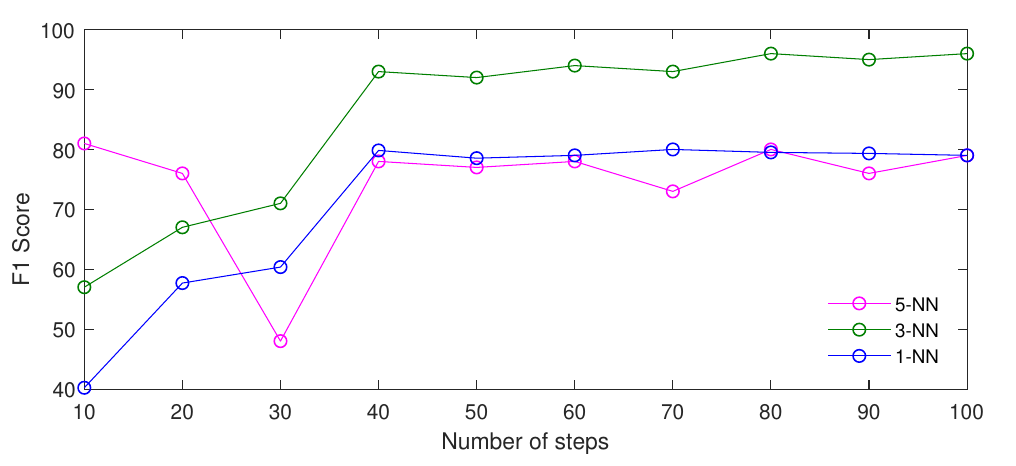}
    \caption{}
    \end{subfigure}    
    \begin{subfigure}{0.49\textwidth}
    \includegraphics[width = 7.5cm,height = 4cm]{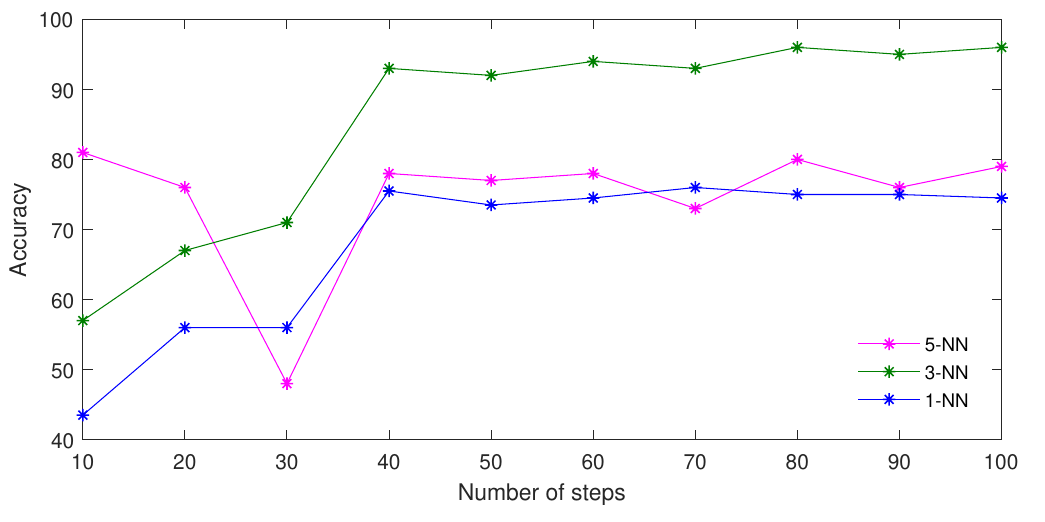}
    \caption{}
    \end{subfigure}      
    \begin{subfigure}{0.49\textwidth}
    \includegraphics[width = 7.4cm,height = 3.9cm]{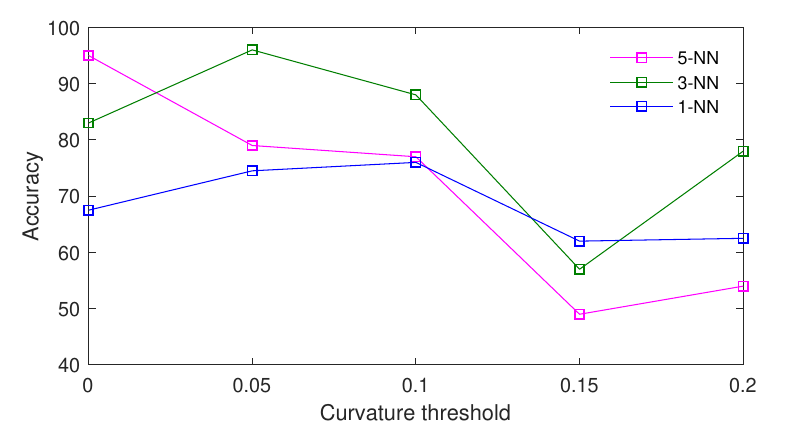}
    \caption{}
    \end{subfigure} 
\caption{\textcolor{black}{Performance evaluation of the proposed method. (a-e): Performance criteria versus the number of steps in Ricci energy optimization and parameter K in KNN. (f): Accuracy criteria versus curvature threshold and parameter K in KNN.}}
\label{figPerfoVsSteps}
\end{figure*}

\begin{figure*}[!h]
\begin{center}
\includegraphics[width=1\textwidth]{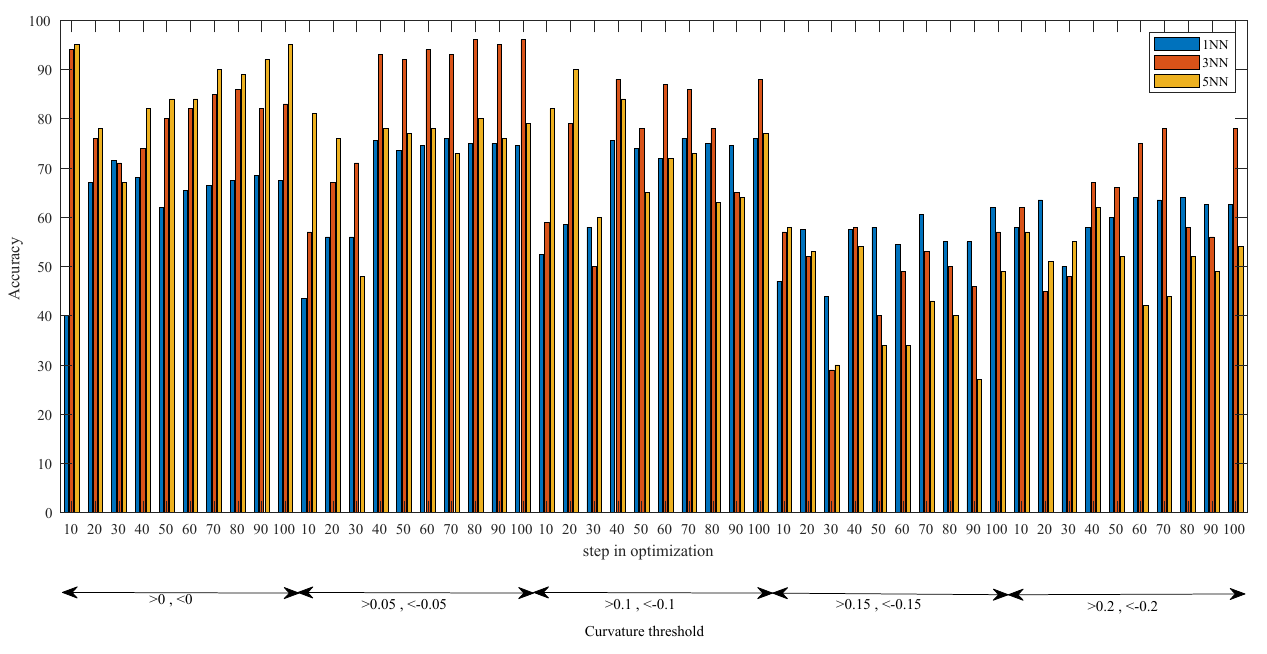}
\caption{\textcolor{black}{Effect of the number of steps in Ricci energy optimization and curvature threshold in selected vertices and K in KNN on the classification performance of the proposed method.}}
\label{figAccuracyVsThreshSteps}
\end{center}
\end{figure*}

For our classification purpose, we used the RBF kernel method to diagnose Alzheimer's disease as described in Section \ref{3Dclassify}. We performed a leaving-one-out cross-validation with the KNN classifier. The proposed method was evaluated with different K in KNN and selected stages in the Ricci flow optimization. The result can be seen in Table \ref{tbl:classify}, Figure \ref{figPerfoVsSteps}, and Figure \ref{figAccuracyVsThreshSteps}. Vertex curvature less than -0.05 and greater than +0.05 with K equal to 3 and step number equal to 80 and 100 resulted in better subject matching and accuracy. Additionally, with K equal to 1, sensitivity of 97\% have been achieved in step number equal to 40, 50, and 80. More clearly Figure \ref{figPerfoVsSteps} (a-e) shows the performance criteria versus step number and Figure \ref{figPerfoVsSteps} (f) shows accuracy versus curvature threshold changes, respectively. While the execution time of the proposed method is acceptable, the code has not been optimized. Our primary goal, however, is not efficiency, but rather ensuring high accuracy in terms of classification.
\newpage
\begin{table}[!h]\scriptsize
	\caption{\textcolor{black}{Effect of the number of steps in Ricci energy optimization and parameter K in KNN with curvature threshold ($<$-0.05, $>$+0.05) on the classification performance (AD vs. CN) of the proposed method.}}	
\begin{center}
		\begin{tabular}{ccccccc}		
			\hline			
			\multicolumn{1}{c}{\textbf{Steps}} &
			\multicolumn{1}{c}{\textbf{K}} &
			\multicolumn{1}{c}{\textbf{ACC}} &
			\multicolumn{1}{c}{\textbf{PRE}} &
			\multicolumn{1}{c}{\textbf{SPE}} &
			\multicolumn{1}{c}{\textbf{SEN}} &			
			\multicolumn{1}{c}{\textbf{F1}} \\	
			 &  & (\%) & (\%) & (\%) & (\%)  & (\%)	\\			
			\hline						
			10 & 1 & 43.5 &	42.7 & 49 & 38 & 40.2 \\			
			    & 3 & 57 & 57 & 57 & 57 & 57\\
			    & 5 & 81 & 81 & 81 & 81 & 81\\
			\hline
			20 & 1 & 56 & 55.56 & 52 & 60 & 57.7 \\			
			    & 3 & 67 & 67 & 67 & 67 & 67\\
			    & 5 &  76 & 76 & 76 & 76 & 76 \\
			\hline
			30 & 1 & 56	& 54.9 & 45	& 67 & 60.36 \\			
			    & 3 &  71 & 71 & 71 & 71 & 71\\
			    & 5 &  48 & 48 & 48 & 48 & 48\\
			\hline		
			40 & 1 & 75.5 &	67.83 &	54 & \textbf{97}	& 79.83 \\			
			    & 3 &  93 &  93 & 93 & 93 & 93\\
			    & 5 &  78 & 78 & 78 & 78 & 78\\
			\hline									
			\textbf{50} & 1 & 73.5 & 65.9 &	50 & \textbf{97} &	78.5 \\
				 & 3 & 92  & 92 & 92 & 92 & 92\\
				 & 5  & 77 & 77 & 77 & 77 & 77\\
			\hline 
			60 & 1 & 74.5 &	67.13 &	53 & 96 & 79 \\
				 & 3 & 94 & 94 & 94 & 94 & 94\\
				 & 5 & 78 & 78 & 78 & 78 & 78\\
			\hline 
			70 & 1 & 76 & 68.57 & 56 &	96	& 80 \\
				 & 3 & 93 & 93 & 93 & 93 & 93\\
				 & 5  & 73 & 73 & 73 & 73 & 73\\			
			\hline			
			\textbf{80} & \textbf{1} & 75 & 67.4 & 53	& \textbf{97} & 79.5 \\
			   & \textbf{3} & \textbf{96} & \textbf{96} & \textbf{96} & \textbf{96} & \textbf{96}\\
			   & 5 & 80 & 80 & 80 & 80 & 80 \\	
			\hline
			90 & 1 & 75 & 67.6 & 54	& 96 &	79.34  \\
			   & 3 & 95 & 95 & 95 & 95 & 95 \\
			   & 5 & 76 & 76 & 76 & 76 & 76\\	
			\hline
			\textbf{100} & 1 & 74.5 &	67.1 & 53 & \textbf{96} & 79 \\
				& \textbf{3} & \textbf{96} & \textbf{96} & \textbf{96} & \textbf{96} & \textbf{96} \\
			    & 5 & 79 & 79 & 79 & 79 & 79\\	
			\hline			
		\end{tabular}
		\end{center}
	\label{tbl:classify}
\end{table}
\newpage
\subsection{Comparison with state-of-the-art methods}
A recent paper by \cite{tanveer2020machine} presented an analysis and comparison of 165 papers for Alzheimer’s Disease (AD) classification on the ADNI dataset. Additionally, the state-of-the-art methods have been compared with ours, as shown in Table \ref{ComparisonTbl}. \textcolor{black}{In \cite{zeng2013teichmuller}, a shape descriptor in Teichmüller space has been presented using Ricci flow; the authors reported an accuracy up to 91.93\% in the AD diagnosis on the ADNI dataset. 
Shi et al. in \cite{shi2019hyperbolic} proposed a hyperbolic Wasserstein distance using Ricci flow and applied it to the categorization of cortical surfaces in the AD/CN problem and obtained a 76.7\% accuracy rate on the ADNI dataset.
In \cite{razib2017structural}, to enable comparison and visualization of the different atlas structures, the author has developed a structural brain mapping that maps the brain surface into a planar convex domain using Tutte's embedding of a new atlas graph and a harmonic map with atlas graph constraints. Shape similarity metrics for AD classification have been computed using the suggested method. The best results with an accuracy of 88.0\% in AD diagnosis on the ADNI dataset have been reported. Acharya et al. \cite{acharya2019automated} suggested a feature extraction method using the Shearlet Transform (ST) to diagnose Alzheimer’s disease. On the ADNI dataset, they attained an accuracy of 94.54\%. In \cite{qin20223d}, a 3D Residual U-Net method with a hybrid attention technique to diagnose Alzheimer's disease on 3D MRI has been proposed. In the AD/CN classification task on the ADNI dataset, an accuracy of 92.68\% was reported. Zhang et al. in \cite{zhang2022diagnosis}, in order to diagnose AD, suggested a deep learning architecture based on sMRI gray matter slices that combined slice region and attention mechanism. They achieved 90\% accuracy in the classification of AD/CN on the ADNI dataset. In \cite{kushol2022addformer}, authors captured the global or long-range relationship of image features by vision transformer architecture. They applied this method to detect Alzheimer’s patients from healthy controls with an accuracy of 88.20\% on the ADNI dataset. Kong et al. \cite{kong2022multi} proposed an image fusion method to fuse MRI with PET images using 3D convolutional neural networks for Alzheimer's disease diagnosis with an accuracy of 93.21\% on the ADNI dataset. }\\
We computed manifold-based classification of covariance matrices computed through Ricci energy optimization in diagnosing Alzheimer's disease by processing hippocampal regions in brain surface data. Our method differs from most existing methods in that it emphasizes local geometries and processes the hippocampal region as a region of interest in AD classification. Based on our statistical results, our conformal structure-based features may offer new structural MRI measures that could improve classification accuracy. However, further validation is required to determine whether this approach is more reliable than other metrics. By way of comparison, our method, with its rich and valuable signatures, achieved an accuracy of 96\%, which is superior to other methods.
\begin{flushleft}
\begin{table}[!h]\scriptsize
	\caption{Comparison of the proposed method with state-of-the-art methods, on the ADNI dataset.}	
	\begin{center}			
		\begin{tabular}{llll}			
		\hline			
		\multicolumn{1}{l}{\textbf{Author}} &
		\multicolumn{1}{l}{\textbf{Year}} &
		\multicolumn{1}{l}{\textbf{Method}} &
		\multicolumn{1}{l}{\textbf{Accuracy}} \\
		\hline
		Zeng et al. \cite{zeng2013teichmuller} & 2013 & Teichmüller shape descriptor + SVM & 91.38 \\				
		Shi et al. \cite{shi2019hyperbolic} & 2019 & hyperbolic Wasserstein distance & 76.7 \\
		Razib \cite{razib2017structural} & 2017 & Tutte's embedding and Harmonic mapping + KNN & 88 \\
		Acharya et al. \cite{acharya2019automated} & 2019 &	Shearlet transform + KNN & 94.54 \\	
		Qin et al. \cite{qin20223d} & 2022 & 3D Residual U-Net model with hybrid attention technique & 92.68\\				
		Zhang et al. \cite{zhang2022diagnosis} & 2022 & sMRI gray matter segments + deep learning & 90 \\
		Kushol et al. \cite{kushol2022addformer} &	2022 & vision Transformer &	88.20 \\
		Kong et al. \cite{kong2022multi}	& 2022 &  fuse MRI with PET images + 3D CNN  & 93.21 \\
		Proposed Method & - &covariance descriptors in Ricci energy optimization + KNN & $\textbf{96}$ \\
		\hline								
		\end{tabular}	
	\end{center}
	\label{ComparisonTbl}
\end{table}
\end{flushleft}
\subsection{\textcolor{black}{An ablation study}}
\subsubsection{\textcolor{black}{Feature-based ablation study}}
\textcolor{black}{In this section, we present the additional experiments conducted to find an effective combination of features and to analyze the contribution of these features to the diagnosis of Alzheimer's disease. As can be seen in Table \ref{tbl:ablation}, the heat kernel feature outperforms the other features, particularly for large steps, and its combination with the other two features is also better than the combination of conformal factor and area distortion. Furthermore, the combination of heat kernel and area distortion is better than the combination of it and conformal factor. }
\begin{table}[!h]\scriptsize
	\caption{\textcolor{black}{Accuracy criteria of the proposed method for analyzing the contribution of features. The heat kernel feature outperformed the other features. Although the area distortion yielded poor accuracy, the combination of this feature and the heat kernel resulted in a relatively good accuracy.}}	
\begin{center}
		\begin{tabular}{ccccccccc}		
			\hline
			\multicolumn{1}{c}{\textbf{Steps}} &
			\multicolumn{1}{c}{\textbf{K}} &									
			\multicolumn{7}{c}{\textbf{Features}} \\
			& & 											
			\multicolumn{1}{c}{\textbf{U}} &			
			\multicolumn{1}{c}{\textbf{AD}} &
			\multicolumn{1}{c}{\textbf{HK}} &			
			\multicolumn{1}{c}{\textbf{[U,AD]}} &
			\multicolumn{1}{c}{\textbf{[U,HK]}} &
			\multicolumn{1}{c}{\textbf{[AD,HK]}} &
			\multicolumn{1}{c}{\textbf{[U,AD,HK]}}  \\
			\hline	
			10  & 1 & 41.5 &	65&	66.5&	60&	50.5&	50	&43.5 \\
				& 3 & 44	& 59 &	10	& 65 &	68 &	63	& 57 \\
				& 5 & 52	& 58	& 28 &	63 &	91 &	80 &	81 \\
				\hline
			20 	& 1 & 47	& 64.5	& 32.5 &	62 &	42	& 51.5	& 56 \\
				& 3 & 56	& 47 &	20	& 53	& 86 &	80 &	67 \\
				& 5 & 56	& 59 &	54 &	60 &	70 &	72 &	76 \\
				\hline
			30  & 1 & 52.5	& 60	& 28	& 67 &	23.5 &	47.5 &	56 \\
				& 3 & 61	& 41	& 4	& 72	& 88	& 69 &	71 \\
				& 5 & 58	& 55	& 7	& 63	& 89	& 58	& 48 \\
				\hline
			40  & 1 & 53.5	& 59.5	& 31	& 69	& 41	& 70.5	& 75.5 \\
				& 3 & 64	& 48	& 80	& 76	& 87	& 90	& 93 \\
				& 5 & 58	& 49	& 70	& 63	& 87	& 75	& 78 \\
				\hline
			50  & 1 & 55.5	& 59	& 30.5	& 68	& 29.5	& 70.5 &	73.5 \\
				& 3 & 61	& 48	& 31 &	77 &	88 &	91 &	92 \\
				& 5 & 58	& 48	& 28	& 64	& 62	& 65 &	77 \\
				\hline
			60  & 1 & 53.5	& 59	& 29.5	& 70	& 29	& 62	& 74.5 \\
				& 3 & 61	& 46	& 83 &	77 &	88 &	90.5 &	94 \\
				& 5 & 58	& 55	& 12	& 64 &	71 &	88	& 78 \\
				\hline
			70  & 1 & 52	& 58.5	& 31.5	& 68	& 45	& 71	& 76 \\
				& 3 & 62	& 43	& 75 &	73 &	87 &	92	 & 93 \\
				& 5 & 58	& 49	& 87	& 65	& 67	& 92 &	73 \\
				\hline
			80  & 1 & 54	& 58.5	& 31.5	& 69	& 35.5	& 71 &	75 \\
				& 3 & 62	& 47	& 84 &	80	& 89	& 93.2 &	\textbf{96} \\
				& 5 & 60	& 52	& 88	& 64 &	79 &	86	& 80 \\
				\hline
			90  & 1 & 52.5	& 58	& 30	& 70.5	& 32.5 &	65.5 &	75 \\
				& 3 & 62	& 48 &	80	& 77 &	90 & 	90 &	95 \\
				& 5 & 58	& 53	& 76	& 64	& 69 &	75 &	76 \\
				\hline
			100 & 1 & 53.5	& 58.5	& 28.5 &	71	& 32	& 65	& 74.5 \\
				& 3 & 62	& 48	& 90	& 77	& 88	& 93	& \textbf{96} \\
				& 5 & 58	& 47	& 89 &	58 &	59 &	90	& 79 \\
				\hline			
		\end{tabular}
\end{center}
	\label{tbl:ablation}
\end{table}
\subsubsection{\textcolor{black}{Evaluation of the proposed method on an imbalanced data set}}
\textcolor{black}{Many real-world data sets naturally exhibit class imbalance, where the number of instances in different classes is not evenly distributed. Evaluating the method on an imbalanced data set provides a more realistic assessment of its performance, helps address biases, and ensures suitability for real-world applications with imbalanced data. Classification algorithms are often biased toward the majority class in an imbalanced data set, leading to high accuracy but poor performance on the minority class. By evaluating the proposed method on an imbalanced data set, we can assess if the method meets the specific requirements of the application, such as correctly detecting rare diseases. Therefore we made an extra experiment on an imbalanced data set containing 85 AD and 100 CN subjects, by randomly removing 15 AD subjects from the main dataset. The classification performances (AD versus CN) are presented in Table \ref{tbl:classifyImbalanced}, as we can see, the result for the minority class is promising and we obtained a high accuracy of 97.84\% for steps equal 80 and 100 in K equal to 3 and specificity and sensitivity of 100\% for all steps in K equal to 3 and 5 on this data set as well.}
\begin{table}[!h]\scriptsize
	\caption{\textcolor{black}{Effect of the number of steps in Ricci energy optimization and parameter K in KNN with curvature threshold ($<$-0.05, $>$+0.05) on the classification performance (AD vs. CN) of the proposed method on imbalanced data set.}	}
\begin{center}
		\begin{tabular}{ccccccc}		
			\hline			
			\multicolumn{1}{c}{\textbf{Steps}} &
			\multicolumn{1}{c}{\textbf{K}} &
			\multicolumn{1}{c}{\textbf{ACC}} &
			\multicolumn{1}{c}{\textbf{PRE}} &
			\multicolumn{1}{c}{\textbf{SPE}} &
			\multicolumn{1}{c}{\textbf{SEN}} &			
			\multicolumn{1}{c}{\textbf{F1}} \\	
			 &  & (\%) & (\%) & (\%) & (\%)  & (\%)	\\			
			\hline						
			10 & 1 & 56.2 &	53.12 &	70 & 40 & 45.63 \\			
			    & 3 & 82.16 &	\textbf{100} &	\textbf{100}	& 61.17	& 75.91 \\
			    & 5 & 91.35 &	\textbf{100}	& \textbf{100} &	81.17 & 89.61 \\
			\hline
			20 & 1 &  53.5 &	49.49	& 50 &	57.64 &	53.26 \\
				& 3 &	84.32	& \textbf{100} &	\textbf{100} &	65.88 &	79.43 \\
				& 5 & 87.56	& \textbf{100}	& \textbf{100} &	72.94	& 84.35 \\
			\hline
			30 & 1 & 53.51 & 49.55 &	43	& 65.88	& 56.56 \\
				& 3 & 87.02 &	\textbf{100} &	\textbf{100} &	71.76 &	83.56 \\
				 & 5 & 76.21 & \textbf{100} &	\textbf{100}	 & 48.23 &	65.07 \\
			\hline		
			40 & 1 & 50.27	& 47.95	& 11	& 96.47	& 64.06 \\
				& 3 & 96.21	& \textbf{100}	& \textbf{100}	& 91.76 &	95.70 \\
				& 5 & 88.64	& \textbf{100} &	\textbf{100} &	75.29 &	85.90 \\
			\hline									
			50 & 1 & 50.81 &	48.23 &	12 &	96.47	& 64.31 \\
				& 3 & 95.67 &	\textbf{100} &	\textbf{100} &	90.58 &	95.06 \\
				 & 5  & 87.56	& \textbf{100}	& \textbf{100}	& 72.94	& 84.35  \\
			\hline 
			60 & 1 & 49.72 &	47.64 &	11 &	95.29 &	63.52 \\
				 & 3 & 96.75	& \textbf{100}	& \textbf{100}	& 92.94	& 96.34 \\
				& 5 & 88.64	& \textbf{100}	& \textbf{100}	& 75.29	& 85.90 \\				
			\hline 
			70 & 1 & 49.72 &	47.64	& 11	& 95.29 &	63.52 \\
				& 3 &  96.21	& \textbf{100}	& \textbf{100} &	91.76 &	95.70 \\
				& 5  & 85.94	& \textbf{100} &	\textbf{100}	& 69.41 & 	81.94 \\				 			
			\hline			
			80 & 1 & 50.27	& 47.95	& 11	& 96.47 & 	64.06 \\
				& 3 & \textbf{97.83} &	\textbf{100} &	\textbf{100} &	95.29 &	97.59 \\
				& 5 & 89.18	& \textbf{100}	& \textbf{100}	& 76.47	& 86.66 \\ 		    	
			\hline
			90 & 1 & 49.72 &	47.64 &	11 &	95.29 &	63.52 \\
				& 3 & 97.29	& \textbf{100} &	\textbf{100} &	94.11	& 96.96 \\
				& 5 & 87.02 &	\textbf{100}	& \textbf{100} &	71.76 &	83.56 \\
			\hline
			100 & 1 & 50.81 &	48.21 &	13	& 95.29	& 64.03 \\
				& 3 & \textbf{97.83} &	\textbf{100}	& \textbf{100} &	95.29 &	97.59 \\
				& 5 & 88.64	& \textbf{100}	& \textbf{100}	& 75.29 & 	85.90\\
			\hline			
		\end{tabular}
		\end{center}
	\label{tbl:classifyImbalanced}
\end{table}
\newpage
\subsubsection{Stability to Noise}
\begin{figure}[!h]
    \centering
    \begin{subfigure}{0.49\textwidth}
    \includegraphics[width = 6.5cm,height = 4cm]{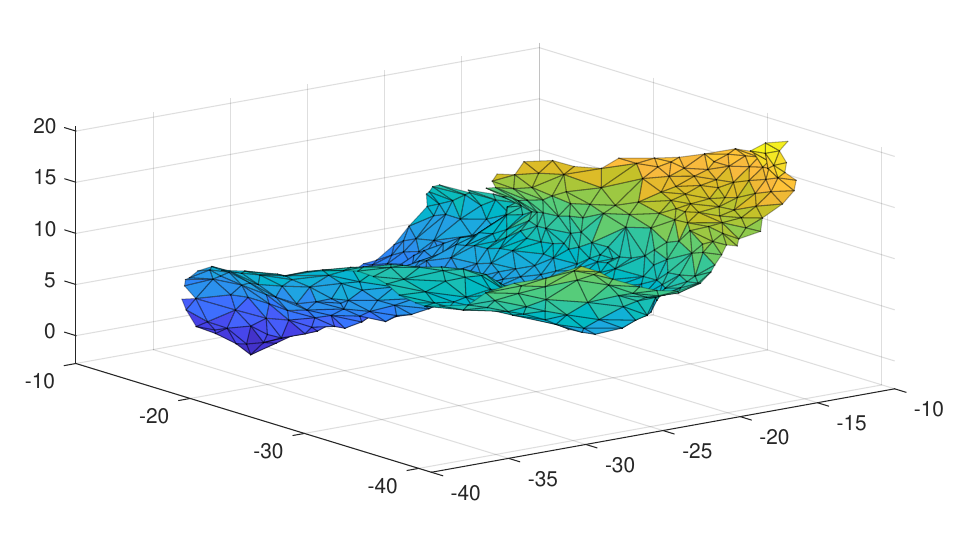}
    \caption{}
    \end{subfigure}
    \begin{subfigure}{0.49\textwidth}
    \includegraphics[width = 6.5cm,height = 4cm]{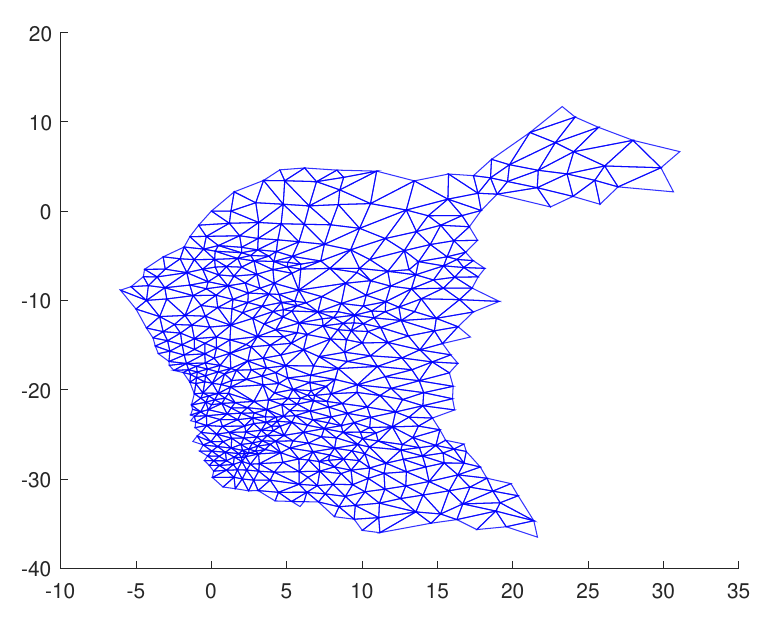}
    \caption{}
    \end{subfigure}
    \begin{subfigure}{0.49\textwidth}
    \includegraphics[width = 6.5cm,height = 4cm]{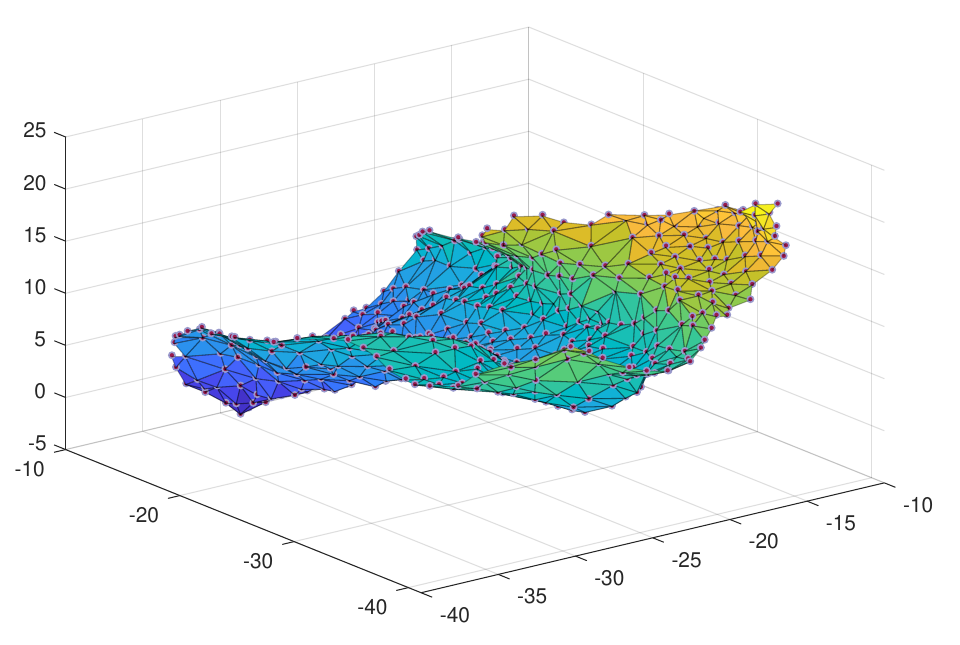}
    \caption{}
    \end{subfigure}
    \begin{subfigure}{0.49\textwidth}
    \includegraphics[width = 6.5cm,height = 4cm]{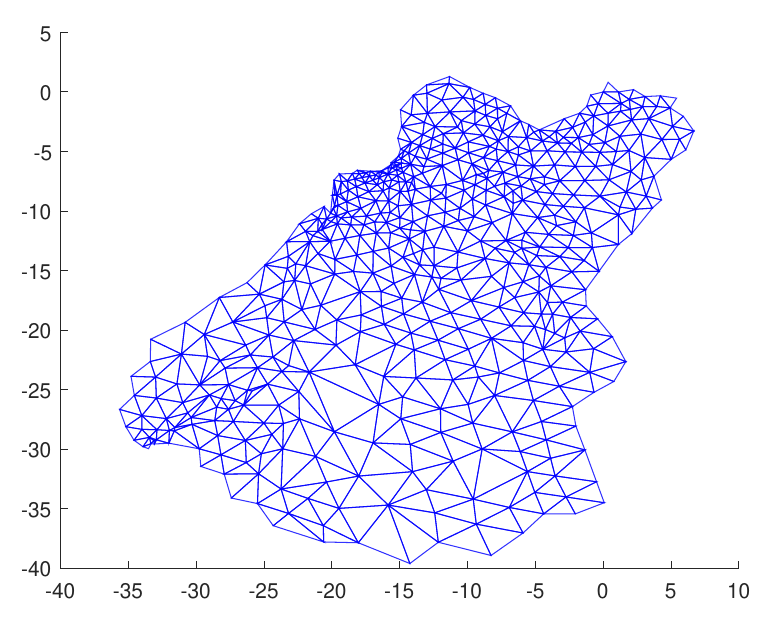}
    \caption{}
    \end{subfigure}  
    \begin{subfigure}{0.49\textwidth}
    \includegraphics[width = 6.5cm,height = 4cm]{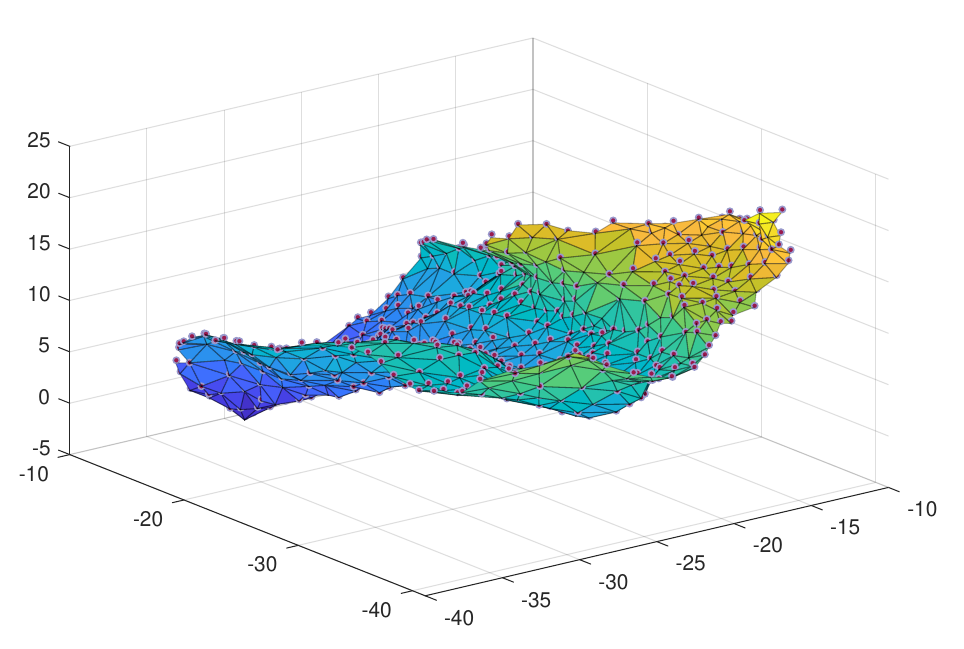}
    \caption{}
    \end{subfigure} 
    \begin{subfigure}{0.49\textwidth}
    \includegraphics[width = 6.5cm,height = 4cm]{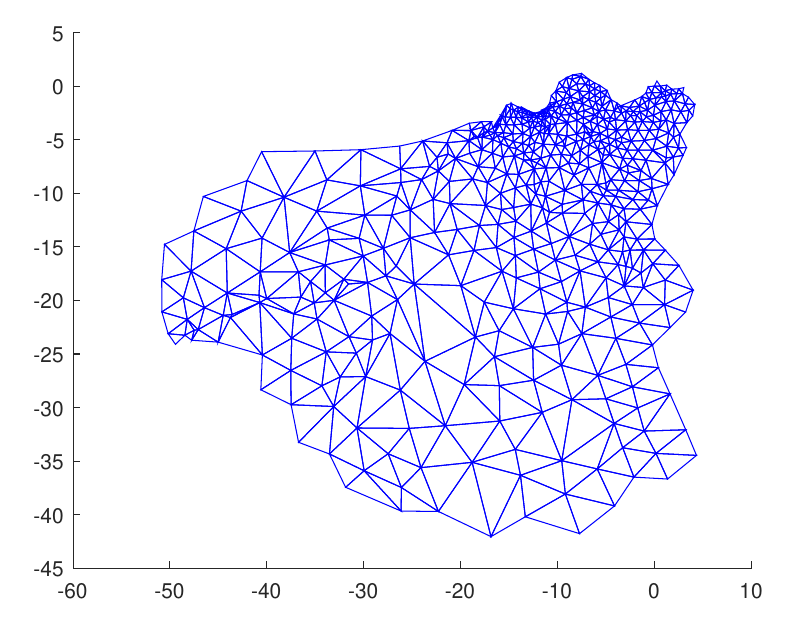}
    \caption{}
    \end{subfigure}  
    \begin{subfigure}{0.49\textwidth}
    \includegraphics[width = 6.5cm,height = 4cm]{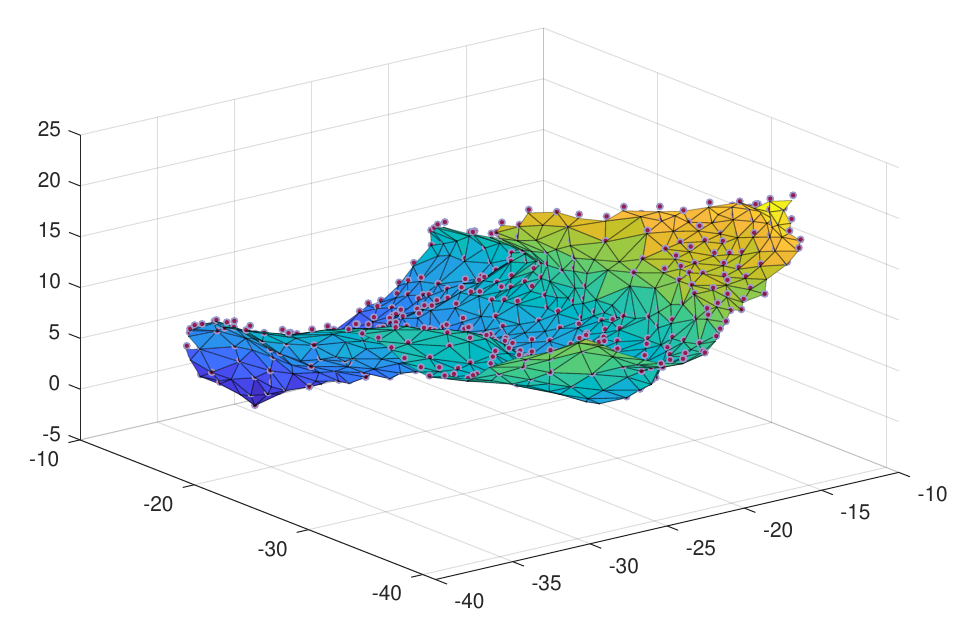}
    \caption{}
    \end{subfigure} 
    \begin{subfigure}{0.49\textwidth}
    \includegraphics[width = 6.5cm,height = 4cm]{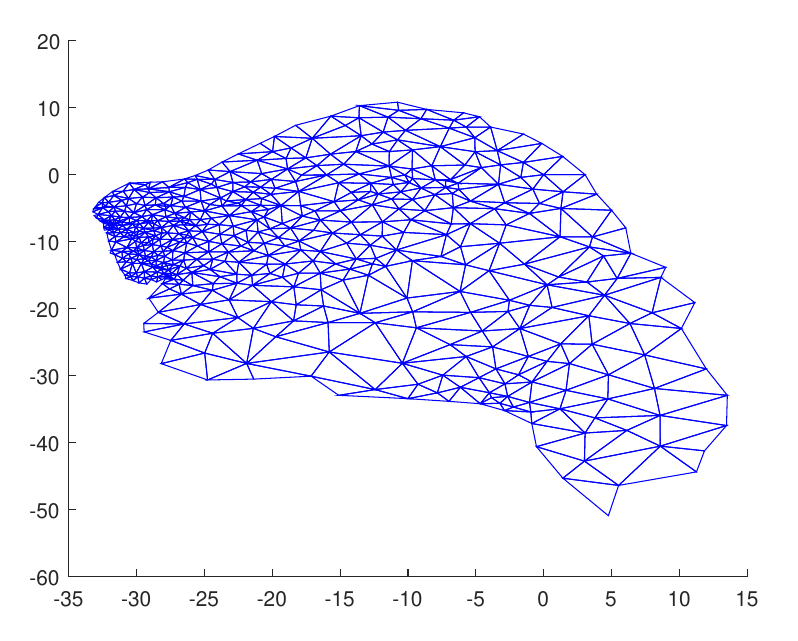}
    \caption{}
    \end{subfigure} 
    \caption{\textcolor{black}{(a): Triangulation mesh of a clean AD subject, (b): Embedding of (a) in the plane after applying Ricci energy optimization. (c): Noisy triangulation mesh of an AD subject in the presence of Gaussian noise with mean equal to 0 and standard deviation equal to 0.1, (d): Embedding of (c) in the plane after applying Ricci energy optimization. (e): Noisy triangulation mesh of an AD subject in the presence of Gaussian noise with mean equal to 0 and standard deviation equal to 0.2, (f): Embedding of (e) in the plane after applying Ricci energy optimization. (g): Noisy triangulation mesh of an AD subject in the presence of Gaussian noise with mean equal to 0 and standard deviation equal to 0.2, (h): Embedding of (g) in the plane after applying Ricci energy optimization.}}
    \label{noisySurf}    
\end{figure}

Geometric noise is any distortion or deviation from the idealized geometric structure of a surface. It can be caused by various factors, including imperfections in the surface itself, changes in the environment, and the effects of other objects in the vicinity. The use of geometric noise in the proposed work allows for a smoother, more accurate solution to the surface Ricci flow, as it can help to reduce the amount of distortion in the resulting surface. \textcolor{black}{Random noise values using a Gaussian distribution with mean equal to 0 and standard deviations equal to 0.1, 0.2 and 0.3 were generated. Although the triangulation mesh and therefore mainly parameters such as metric, angle and curvature may vary due to injected noises, the extracted features are consistent characteristics and do not alter the classification result. An AD subject in the presence of these Gaussian noises and embedding of it in the plane can be seen in Figure \ref{noisySurf}.}
\subsubsection{\textcolor{black}{Limitations}}
\textcolor{black}{Let us mention that our current Ricci flow setting is restricted to triangular meshes. This restricts the applicability of our Ricci flow-based classification method to other types of data structures. Another challenge in this context of application is the limited access to 3D brain surface databases, which restricts the availability of data for analysis and modeling purposes.}
\section{Conclusion and future work}\label{sec13}
In this paper, we proposed a 3D surface classification method based on covariance descriptors in the optimization step of Ricci flow. The method yields an accuracy of 96\% in distinguishing patients with Alzheimer’s disease from normal subjects, demonstrating its superiority compared to state-of-the-art methods. Furthermore, the signatures effectively indicate surface information and can be used as signatures in neuroimaging studies, \textcolor{black}{such as intermediate stages (SMC, EMCI, and LMCI) of Alzheimer's disease. 
Our proposed methodology can be applied to all those intermediate stages of Alzheimer's disease. We didn’t perform experiments on those classes, since such multi-class classification would require much more data, in order to have a meaningful number of samples per class. This would be challenging in our context since the preprocess (reconstructing  cortical surface using freesurface) is time-consuming.} The same applies to Schizophrenia patterns, considering additional features such as cortical thickness, \textcolor{black}{which remains an interesting avenue for future research.}
Additionally, this approach offers potential applicability to other surfaces beyond brain cerebral cortices. For instance, interestingly,  our proposed classification strategy may find utility in recognizing facial expressions, based on recent related works \cite{liu20234d, Lei2021}.
\clearpage 
\newpage
\bibliographystyle{IEEEtranN}
\bibliography{sample}
\end{document}